\documentclass[fleqn,usenatbib]{mnras}

\usepackage{newtxtext,newtxmath}
\usepackage{multirow}
\usepackage{array}  
\newcolumntype{C}[1]{>{\centering\arraybackslash}m{#1}}

\usepackage[T1]{fontenc}

\newcommand{\ha}{\hbox{H$\alpha$}}
\newcommand{\hb}{\hbox{H$\beta$}}

\newcommand{\hii}{\hbox{H\,{\sc ii}}}

\newcommand{\oii}{\hbox{[O\,{\sc ii}]}}
\newcommand{\oiii}{\hbox{[O\,{\sc iii}]}}
\newcommand{\oiiisemi}{\hbox{O\,{\sc iii}]}}
\newcommand{\cii}{\hbox{[C\,{\sc ii}]}}
\newcommand{\ciiisemi}{\hbox{C\,{\sc iii}]}}
\newcommand{\nii}{\hbox{[N\,{\sc ii}]}}
\newcommand{\niii}{\hbox{[N\,{\sc iii}]}}

\newcommand{\neiii}{\hbox{[Ne\,{\sc iii}]}}

\newcommand{\heii}{\hbox{He\,{\sc ii}}}

\DeclareRobustCommand{\VAN}[3]{#2}
\let\VANthebibliography\thebibliography
\def\thebibliography{\DeclareRobustCommand{\VAN}[3]{##3}\VANthebibliography}

\usepackage{graphicx}	
\usepackage{amsmath}	

\title[ASTRID Emission Lines]{Theoretical emission lines and metallicity calibrations of H II regions in ASTRID simulation}

\author[Yao Yao et al.]{
Yao Yao,$^{1}$\thanks{E-mail: yao.yao@anu.edu.au}
Kathryn Grasha,$^{1}$
Stuart Wyithe,$^{1}$
Enci Wang,$^{2}$
Nianyi Chen,$^{4}$
Patrick Lachance,$^{3}$
\newauthor
Tiziana Di Matteo,$^{3}$
and Yihao Zhou$^{3}$
\\
$^{1}$Research School of Astronomy and Astrophysics, Australian National University, Weston Creek, Canberra 2611, Australia\\
$^{2}$Deep Space Exploration Laboratory / Department of Astronomy, University of Science and Technology of China, Hefei 230026, P. R. China\\
$^{3}$McWilliams Center for Cosmology,Department of Physics, Carnegie Mellon University, Pittsburgh, PA 15213, USA\\
$^{4}$School of Natural Sciences, Institute for Advanced Study, Princeton, NJ 08540, USA
}

\date{Accepted XXX. Received YYY; in original form ZZZ}

\pubyear{\the\year{}}

\begin{document}
\label{firstpage}
\pagerange{\pageref{firstpage}--\pageref{lastpage}}
\maketitle

\begin{abstract}
We present a theoretical framework to derive redshift-dependent metallicity calibrations for galaxies at $z$=2-7. The ionization parameter ($U$) and gas pressure ($P$) in our approach are not assumed, but are predicted self-consistently. By combining the ASTRID cosmological simulation with stellar population synthesis (SPS) and MAPPINGS V photoionization modeling, we evolve young star clusters under an analytic wind-driven bubble model. This directly couples stellar feedback to the local ISM density, allowing \hii{} region properties to emerge from the underlying physics rather than being treated as free parameters.
The emission-line predictions are validated against observed star-formation rate indicators (deviation <0.05 dex) and the \oiii{} luminosity function. We derive calibrations for common optical (e.g. R23, O3N2, N2, O32) and UV (e.g. C3O3, N3O3) diagnostics. We find significant redshift evolution in these relations, driven primarily by changing ionization conditions. A Bayesian analysis quantifies calibration performance under varying signal-to-noise, enabling diagnostic recommendations as a function of redshift and data quality. The R23 calibration performs well at all redshifts with minimal error in our model, while nitrogen- and carbon-based calibrations are highly sensitive to the abundance enrichment process and should be used with caution. These results provide a practical framework for interpreting JWST spectroscopy and tracing chemical evolution from cosmic noon to the epoch of reionization.

\end{abstract}

\begin{keywords}
Metallicity -- H II regions -- High-redshift galaxies
\end{keywords}



\section{Introduction}

The gas-phase metallicity ($Z_g$) of galaxies most commonly traced by the oxygen abundance (O/H) serves as a key diagnostic of galaxy evolution, reflecting the complex interplay between star formation, gas inflows, and outflows \citep[e.g.,][]{Lilly2013,Maiolino2019,Wang2021,Wang2022,Chen2025,Yu2025,Lyu2026}. In the local universe, metallicity is observed to correlate strongly with stellar mass, forming a well-defined mass-metallicity relation \citep[MZR;][]{Lequeux1979,Tremonti2004,Ma2024,Jia2025}. This relation has been shown to evolve with redshift, with galaxies at earlier cosmic epochs exhibiting lower metallicities at fixed stellar mass \citep[e.g.,][]{Zahid2014,Gao2018,Sanders2021,Yao2022a}. Understanding the evolution of the MZR and its scatter is crucial for constraining models of baryon cycling and feedback processes across cosmic time. Measurements of $Z_g$ are typically derived from emission-line ratios \citep{Kewley2019}, but these measurements have faced some difficulties in the high-redshift universe due to some observational or empirical issues, thus posing a fundamental challenge to our understanding of galaxy formation and evolution.

One of the most reliable methods to determine gas-phase metallicity is the ``direct-T$_e$ method'', which relies on the detection of faint auroral lines such as \oiii$\lambda$4363, \nii$\lambda$5755, and \oiiisemi$\lambda$1666 to derive electron temperatures (T$_e$) and then compute ionic abundances \citep[e.g.,][]{Osterbrock2006,Izotov2006,Pilyugin2010,Mingozzi2022}. However, due to the intrinsic faintness of these lines, their detection has historically been limited to nearby galaxies or stacked spectra \citep[e.g.,][]{Andrews2013,Gao2017,Bian2018,Yao2022,Zou2024}. Before the launch of the James Webb Space Telescope (JWST), only a few dozen galaxies at $z$>1 had auroral-line detections, often biased toward systems with high specific star formation rates \citep[e.g.,][]{Sanders2020,Nakajima2022}.

Due to the absence of auroral lines, metallicity estimates in most of galaxies have to rely on strong-line calibrations--empirical or theoretical relations between bright emission-line ratios (e.g., R23, N2, O3N2, N2O2, N2S2\ha) and oxygen abundance. These calibrations are typically derived from local \hii{} regions or low-redshift galaxy samples \citep[e.g.,][]{Pettini2004,Maiolino2008,Marino2013,Curti2017} or photoionization models \citep[e.g.,][]{McGaugh1991,Kewley2002,PerezMontero2014,Dopita2016}, limited by the physical condition of the interstellar medium (ISM) in the local universe. However, growing evidence especially JWST's observation suggests that the ISM conditions in high-redshift galaxies such as ionization parameter ($U$), electron density, and hardness of the ionizing spectrum can differ significantly from those in typical local galaxies \citep{Steidel2014,Kewley2015,Kaasinen2017,Kaasinen2018,Sanders2020,Abdurrouf2024,Topping2025}. Therefore, applying local strong-line calibrations to high-redshift galaxies may introduce significant systematic biases, which can lead to underestimated metallicities and misinterpretations of chemical evolution \citep{Bian2018,Hirschmann2023}.

Thanks to the high sensitivity and wide wavelength range of JWST, it has become possible to measure the metallicity of high-redshift galaxies up to $z\sim9$ through the direct-T$_e$ method \citep[e.g.,][]{Curti2023,Nakajima2023,Laseter2024,Sanders2024,Chakraborty2025}. They show significant deviations from local relations, but the samples are still small and limited. Therefore, it is necessary to create a theoretical framework that can span the vast parameter space of the high-redshift ISM, which can avoid selection bias and observational limitations.



Cosmological simulations coupled with photoionization models provide a powerful tool for predicting emission-line ratios across a wide range of redshifts. \cite{Hirschmann2023,Hirschmann2023a} utilized IllustrisTNG simulations coupled with a library of emission-line templates including star formation, AGN, and shocks, while \cite{Garg2024} employed SIMBA simulations integrated with \texttt{CLOUDY} to account for \hii{} regions, post-AGB stars, and diffuse ionized gas. While these studies successfully demonstrate that strong-line ratios evolve with redshift, their predictive power is fundamentally limited by their reliance on pre-defined empirical scaling relations, such as fixed gas densities, empirical N/O--O/H relationship curve, assumed $\log U$--$\log(\rm O/H)$ correlation, and an assumed equivalence between stellar and gas-phase metallicities. These models often ``lock in'' local physics, potentially masking the very environmental diversities that define the high-redshift ISM. Consequently, these frameworks are not fully self-consistent, as the resulting line ratios are largely a reflection of the input empirical constraints rather than an emergent property of the simulated galaxy evolution. Despite using high-redshift simulation data, they remain, in essence, variants of local strong-line calibrations that fail to fundamentally address the issue of unique environmental conditions of the early universe.

In this paper, we fully exploit the information advantages provided by the ASTRID cosmological simulation \citep{Bird2022,Ni2022,Ni2025} to derive the nebular emission of galaxies at $z$=2-7 by coupling the simulated star and gas properties with a theoretical \hii{} region model and a photoionization code. Unlike previous works, our approach allows the ionization parameter ($U$) and gas pressure to be determined naturally from the local galactic environment. This approach allows us to derive physically motivated, self-consistent redshift-dependent calibrations grounded in cosmological simulations for strong optical and UV line ratios that are specifically tailored to the evolving physical conditions of high-redshift galaxies. By bridging the gap between small-scale nebular physics and large-scale galaxy assembly, our self-consistent model can provide a critical and timely tool to interpret the growing archive of JWST spectra and decode the extreme condition from the epoch of reionization to cosmic noon. Throughout this paper, we adopt a flat $\Lambda$CDM cosmology with H$_0$ = 67.74 km~s$^{-1}$~Mpc$^{-1}$, $\Omega_0$ = 0.3089, and $\Omega_\Lambda$ = 0.6911.


\section{Theoretical framework}

\subsection{The ASTRID simulation}

We base our analysis on the ASTRID cosmological hydrodynamic simulation, which is run with the MP-Gadget code \citep{Springel2005,Feng2018,Bird2022,Ni2022}
and designed to model galaxy formation and evolution from the epoch of reionization to the present universe. The simulation evolves a comoving volume of (250 $h^{-1}$ Mpc)$^3$ from $z$=99 to now, with $2\times5500^3$ particles, yielding a dark matter particle mass resolution of $6.74\times10^6~h^{-1}$ M$_\odot$ and an initial gas particle mass of $1.27\times10^6~h^{-1}$ M$_\odot$. Compared to its predecessor BlueTides \citep{Feng2016}, ASTRID features a factor of $\sim$2 improvement in mass resolution and extends the redshift range to now. Compared with other simulations, such as IllustrisTNG \citep{Nelson2018,Springel2018} and EAGLE \citep{Schaye2015}, it possesses a larger volume than TNG300 (205 Mpc $h^{-1}$ box), while maintaining a mass resolution comparable to the higher-resolution TNG100 ($7.5\times10^6$ M$_\odot$ in 75 Mpc $h^{-1}$ box) and EAGLE ($9.7\times10^6$ M$_\odot$ in 70 Mpc $h^{-1}$ box).

ASTRID incorporates a comprehensive physics model, including: Hydrodynamics solved with a pressure-entropy formulation of smoothed particle hydrodynamics \citep[pSPH;][]{Hopkins2013}; Radiative cooling and star formation with a molecular hydrogen-based prescription; Supernova feedback implemented via kinetic winds; Metal enrichment tracing nine individual species (H, He, C, N, O, Ne, Mg, Si, Fe); Inhomogeneous hydrogen and helium reionization with spatially and temporally varying UV backgrounds; Black hole formation, growth, and dynamical friction-driven mergers; Massive neutrinos and the relative velocity between baryons and dark matter.

This simulation has been validated against a range of observational data from low to high redshift, showing broad consistency with existing constraints, covering fields such as galaxy morphology \citep{LaChance2025}, high-$z$ galaxy stellar mass function and UV luminosity function \citep{Bird2022}, black holes \citep{Ni2022,Ni2025,LaChance2026,Zhou2026}, and gravitational wave \citep{Chen2025a,Zhou2025}. Its large volume and high resolution make it particularly suited for studying the statistical properties of high-redshift galaxies and their emission-line signatures.

\subsection{Sample selection} \label{subsubsec:sample}

\begin{figure*}
	\includegraphics[width=\textwidth]{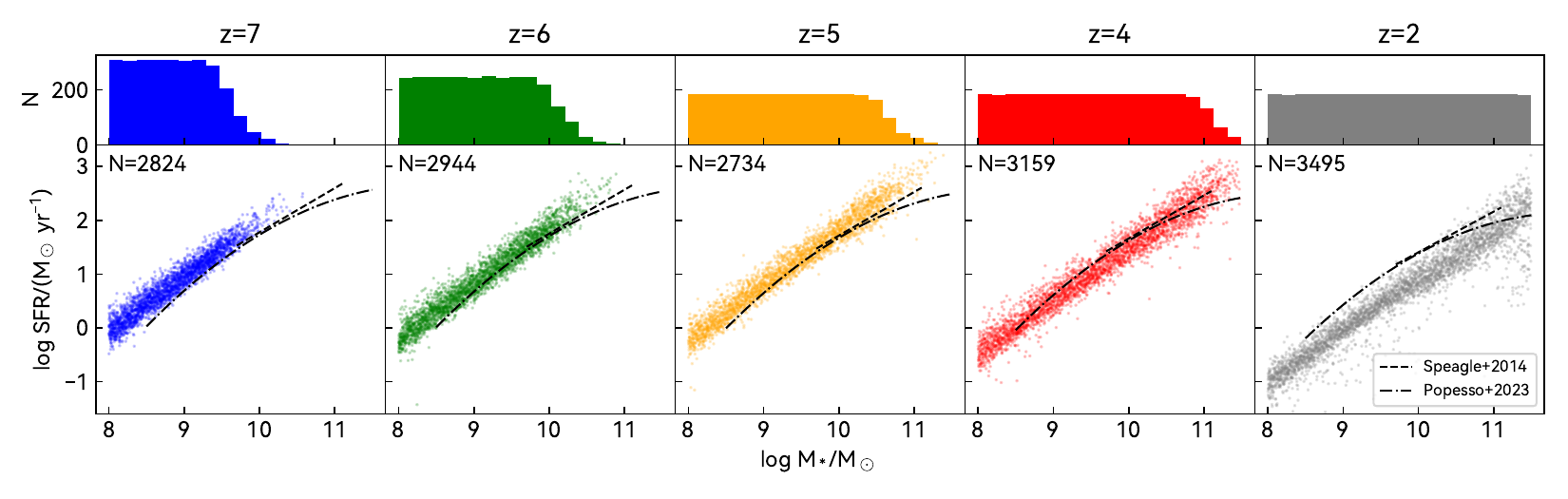}
    \caption{The SFMS of our sample galaxies, top panels are the stellar mass (log(M$_*$/M$_\odot$)) distribution of our sample. Dashed \citep{Speagle2014} and dot-dashed \citep{Popesso2023} curves are the results fitted at corresponding redshift.}
    \label{fig:sfms}
\end{figure*}

Particles in ASTRID simulation are collected into halos using a friends of friends algorithm \citep{Davis1985} and then post processed into subhalos using SUBFIND software \citep{Springel2001}. These subhalos represent the halo substructures associated with galaxies within the simulation. We randomly select a sample of stellar masses with a uniform distribution from subhalo's catalog in a range of log M$_*$ = 8-11.5 at five snapshots (31, 47, 107, 147, 348, $z$=7, 6, 5, 4, 2). Since the stellar mass function is characterized by a high number of low-mass galaxies and a low number of high-mass galaxies, uniform mass sampling would result in incompleteness at the low-mass end and a decrease at the high-mass end. Fig. \ref{fig:sfms} shows the star-formation main sequence \citep[SFMS, e.g.,][]{Salim2007,Whitaker2014,Popesso2023} and stellar mass distribution of our sample. The star formation rates (SFRs) are derived directly from the catalog of ASTRID.

Each star particle in ASTRID is treated as a simple stellar population (SSP), defined by its specific stellar mass ($M_*$), age ($t_*$), and metallicity ($Z_*$). Since we are primarily interested in the nebular emission from \hii{} regions, we only select young star particles ($t_*=0.1-12$ Myr) that host massive O and B stars capable of producing ionizing photons ($\lambda$<912\AA).

\subsection{Ionization source and ISM environment} \label{subsec:ionization_env}

\subsubsection{Star properties} \label{subsubsec:star}

We use \texttt{Starburst99} \citep{Leitherer1999,Vazquez2005,Leitherer2010} to synthesize the integrated spectra, ionizing photon production rates ($Q$), mechanical luminosities ($L_{\rm mech}$), and stellar type distributions for stellar populations across a grid of ages and metallicities.

We generate four time-dependent grids spanning 0.1 to 12 Myr for four metallicities $Z_*$=0.001,0.008,0.02,0.04. These grids include:
\begin{enumerate}
    \item The ratio of the current remaining stellar mass to the initial stellar mass ($M_*(t_*)/M_{*,0}$);
    \item The stellar spectral energy distribution (SED, $L_\lambda(\lambda,t_*)$) per $M_{*,0}$;
    \item $L_{\rm mech}(t_*)$ per $M_{*,0}$,
    \item The number of O-type stars ($n_{\rm O}(t_*)$) per $M_{*,0}$.
\end{enumerate}
We interpolate these grids linearly in age and logarithmically in metallicity using the \texttt{RegularGridInterpolator} from \texttt{scipy}. For a given star particle with properties $M_*$, $t_*$, $Z_*$, we first obtain the initial mass $M_{*,0}$ from the first interpolated grid. The SED, $L_{\rm mech}$ and $n_{\rm O}$ for the particle are calculated by multiplying the corresponding per-initial-mass quantities by $M_{*,0}$. Actually, due to the splitting of star particles, $M_*$ used here needs to be redefined, which will be described in Sect. \ref{subsubsec:split}. The ionizing photon production rate $Q$ can be obtained by integrating the SED below 912\AA.

In our configuration of \texttt{Starburst99}, we assume a \cite{Kroupa2001} initial mass function (IMF) with a high-mass cut of 100 M$_\odot$ and Geneva high evolutionary track. We note that ASTRID simulation uses \cite{Chabrier2003} IMF for chemical enrichment. Although the Chabrier and Kroupa IMFs are the same at the high-mass end relevant for ionizing photon production, their low-mass ends differ, leading to a slight offset when normalizing to unit SFR. Specifically, for the same SFR, the Chabrier IMF will produce $\sim$6\% more ionizing photons than the Kroupa IMF \citep{Speagle2014}, due to its lower total stellar mass at fixed high-mass normalization. Therefore, this difference has a negligible impact on our results regarding \hii{} region emission.

\subsubsection{Gas properties} \label{subsubsec:gas}

Some observations have already found that the gas density will evolve with redshift. High-redshift galaxies tend to have dense gas in \hii{} regions \citep[e.g.,][]{Abdurrouf2024,Topping2025}. Therefore, the assumption of a constant gas density could not be physical in high-redshift galaxies. Here, for the \hii{} region of each young star particle, we compute the local gas density and elemental abundances by considering its 113 nearest gas particles (this is the SPH neighbor count from ASTRID's smoothing kernel). The contribution of each gas particle is weighted by the quintic spline kernel \cite[][Eq. 8]{Price2012}, where the smoothing length $h$ is set to $l_S/3$. $l_S$ is the gas particle's smoothing length derived from the ASTRID catalog.

Given that the gas particles in the simulation are multiphase, we use only the mass of the neutral gas component within each particle for the density calculation. The resulting density, denoted as $\rho_0$, serves as the initial local ISM density for the wind-driven bubble model described in Sect. \ref{subsec:wind_drived}. We find that $\rho_0$ at $z=7$ is $\sim$0.7 dex higher than $z=2$, which is consistent with the result of \cite{Topping2025}.

ASTRID tracks seven metal species (C, N, O, Ne, Mg, Si, Fe) but does not include sulfur. To estimate the sulfur abundance, we scale the silicon abundance relative to the solar values: S=S$_\odot\times$Si/Si$_\odot$, where the solar abundances are S$_\odot= 1.318\times10^{-5}$ and Si$_\odot = 3.236\times10^{-5}$ \citep{Asplund2009}. Since the abundance of sulfur is estimated, the strong line ratios of our metallicity calibrations do not include sulfur emission lines. The abundance of each galaxy used for calibrations is the \hb{} luminosity weighted average of all its \hii{} regions.

\subsubsection{Splitting the star particles} \label{subsubsec:split}

The typical mass of a star particle in ASTRID is $\sim5\times10^5$ M$_\odot$. Modeling a single \hii{} region around such a massive particle would lead to an overestimation of the ionizing photons $Q$ and the ionization parameter $U$ \citep{Garg2024}. To address this resolution limitation, we follow a subgrid approach similar to \cite{Garg2024} to split each star particle into a collection of unresolved small clusters with individual \hii{} regions but share the same $Z_*$ and gas environment. Unlike \cite{Garg2024}, we only split and redistribute the $M_{*,0}$, without redistributing the $t_*$. We use the same method as \cite{Garg2024} to test the effect of age distribution and find that if age distribution is considered, the number of ionizing photons ($Q$) will differ by $\sim$0.05 dex, which is negligible. Therefore, to reduce computation time, we do not continue to split star particles by $t_*$.

We redistribute $M_*$ of star particles according to a cluster mass function (CMF) with a power-law index $\beta=-2$ \citep{Krumholz2019}, spanning the range $M_{\rm min}-$400 000 M$_\odot$ using the method of inverse transform sampling. The lower mass limit $M_{\rm min}$ is typically set to 2 000 M$_\odot$, but it would be dynamically increased, if necessary, to ensure that each sampled cluster contains at least one O-type star based on the precomputed $n_{\rm O}$ per $M_{*,0}$. This is to prevent old \hii{} regions from extinguishing too early. In effect, this is equivalent to the overlapping and merging of \hii{} shells in small clusters as they expand with age in our wind-driven bubble model (see Sect. \ref{subsec:wind_drived}). We also change the $M_{\rm min}$, $M_{\rm max}$ and $\beta$ of the CMF, and find that the variations of the derived emission line ratios (see Sect. \ref{subsec:photo_model}) are low when we change $M_{\rm min}$, $M_{\rm max}$ in $\sim$0.3 dex and change $\beta$ from -1.8 to -2.2 (see Appendix \ref{apdx:split}).

In principle, this stochastic sampling would generate a large number of low-mass clusters with different $Q$ and $L_{\rm mech}$, significantly increasing the computational cost of photoionization calculations. Therefore, we group the sampled clusters into $N_{\rm bin}=4$ mass bins and replace the individual masses in each bin with the bin's average mass. The total emission-line luminosity for star particle $i$ is then given by:
\begin{equation}
    L_i=\sum_{j=0}^{N_{\rm bin}-1}n_{i,j}L_{i,j},
	\label{eq:l_partical}
\end{equation}
where $n_{i,j}$ is the number of small clusters in the mass bin $j$, $L_{i,j}$ is the output luminosity of the photoionization model when $M_*$ is set to the average mass of the mass bin $j$.

\subsection{Wind-driven bubble model} \label{subsec:wind_drived}

The ionization parameter ($U$) of the \hii{} regions at high redshift has been an open question, with some studies linking $U$ to $t_*$ and $Z_*$ \citep[e.g.,][]{Carton2017,Wilkins2020}, while others to SFR \citep[e.g.,][]{Hirschmann2017}. However, the actual factors that determine $U$ are diverse, including the total number of ionizing photons ($Q$), hydrogen density ($n_{\rm H}$), and the distance between the ionized gas and source ($R$). Therefore, we need a model that can physically control the ionization parameters, rather than empirical or approximate relationships.

We adopt the dynamic \hii{} region evolution model presented by \citet{Dopita2006}. Unlike the classical Stromgren sphere and plane-parallel approximation, which assumes a static gas distribution, the wind-driven model couples the physical properties of the ionized gas directly to the time-dependent energy output of the central star cluster. In this framework, for each of our split star particles, the pressure ($P$) and $U$ are physically determined by the interplay between the mechanical luminosity ($L_{\rm mech}$) of the stellar winds/supernovae provided by the SPS and the initial gas density ($\rho_0$) of the ambient ISM provided by the ASTRID gas particles. This process is implemented by controlling the radius ($R$) of the \hii{} shell as follows.

The \hii{} region is treated as a mass-loss blown bubble. The interior is filled largely by hot, shocked stellar wind material at coronal temperatures, while the photoionized \hii{} region exists as a thin shell located between the contact discontinuity and the outer shocked ISM. The expansion of this shell is governed by the equations of conservation of mass, momentum, and energy. Assuming a standard equation of state with $\gamma=5/3$, the time evolution of \hii{} shell radius $R$ can be described by the following equation \citep[][Eq. 11]{Dopita2006}:
\begin{equation}
    \frac{d}{dt}\left[R\frac{d}{dt}(R^3\dot{R})\right]+\frac{9}{2}R^2\dot{R}^3=\frac{3L_{\rm mech}(t)}{2\pi\rho_0},
	\label{eq:hii_r}
\end{equation}
which can be rewritten in the form of:
\begin{equation}
    \dddot{R}(R,\dot{R},\ddot{R}, t)=\frac{3L_{\rm mech}(t)}{2\pi\rho_0}R^{-4}-\frac{27}{2}\dot{R}^3R^{-2}-10\ddot{R}\dot{R}R^{-1}.
	\label{eq:hii_r3}
\end{equation}
We integrate this ordinary differential equation using a standard fourth-order Runge-Kutta method (implemented via \texttt{solve\_ivp} of \texttt{scipy.integrate} package). The solution of Eq. \ref{eq:hii_r3} (consistent with the analytical similarity solution of \citealt{Castor1975}) reveals a scaling relation $R(t)\propto(L_{\rm mech}(t)/\rho_0)^{1/5}$. Since mechanical luminosity scales linearly with the cluster mass ($L_{\rm mech}(t) \propto M_{*,0}$), the radius scales as $R (t)\propto (M_{*,0}/n_{\rm H,0})^{1/5}$, where $n_{\rm H,0} = \rho_0 X / m_p$ is the initial hydrogen number density, and $X=0.76$ is the hydrogen mass fraction.

This scaling allows us to decompose the radius into a component dependent on stellar evolution ($t_*, Z_*$) and a component dependent on the initial conditions ($M_{*,0}, n_{\rm H,0}$):
\begin{equation}
    R(t_*,Z_*,M_{*,0},n_{\rm H,0}) = R_1(t_*, Z_*)\left(\frac{M_{*,0}/M_\odot}{n_{\rm H,0}/{\rm cm}^{-3}}\right)^\frac{1}{5},
	\label{eq:hii_r_new}
\end{equation}
where $R_1(t_*, Z_*)$ is a function in unit of pc that encapsulates the dependence on stellar age and metallicity.
It can be precomputed by numerically integrating Eq. \ref{eq:hii_r} for a reference cluster of $M_{*,0}=1$ M$_\odot$ and $n_{\rm H,0}=1$ cm$^{-3}$ across a grid of $t_*$ and $Z_*$, where the $L_{\rm mech}$ from \texttt{Starburst99} depends on $t_*$ and $Z_*$ and is scaled by $M_{*,0}$, while other parameters can be directly derived from the ASTRID particles (in Sect. \ref{subsec:ionization_env}). This means that $R_1$ can be interpolated following the same procedure as the four star properties in Sect. \ref{subsubsec:star}. Thus, the inner radius $R$ for any arbitrary cluster is then efficiently obtained via Eq. \ref{eq:hii_r_new}.

The pressure of \hii{} region can be derived by Eq. 3 of \cite{Dopita2006}:
\begin{equation}
    P(t)=\frac{7}{(3850\pi)^\frac{2}{5}}\left(\frac{250}{308\pi}\right)^\frac{4}{15}\left(\frac{L_{\rm mech}(t)}{\rho_0}\right)^\frac{2}{3}\frac{\rho_0}{R(t)^\frac{4}{3}},
	\label{eq:hii_p}
\end{equation}
which serves as the input to the following photoionization model.

Although this model provides a self-consistent framework for local starbursts, we should note that there are still some limitations of our model when applied to high-redshift galaxies. Firstly, a single high-redshift \hii{} region may exhibit a density-stratified or multi-phase structure, as demonstrated by the discovery of two-phase ionized gas (dense core vs. diffuse envelope) in systems like COS-2987 \citep{Usui2025}. Such complexities deviate from the single-phase homogeneous medium assumption in the \citet{Dopita2006} model. Furthermore, current studies suggest that $U$ at $z > 6$ may be primarily modulated by the SFR surface density and potential nebular overlap rather than by metallicity-regulated feedback alone \citep{Reddy2023}.

\subsection{Photoionization model} \label{subsec:photo_model}

The resulting set of \hii{} region parameters (metallicity, $U$, pressure) is used as input to the \texttt{MAPPINGS V} 5.2.0\footnote{\url{https://mappings.anu.edu.au/}} photoionization code to predict the luminosity of key optical and UV emission lines.

We use the P6 model (default dust-free pure photoionization) of \texttt{MAPPINGS V} 5.2.0 photoionization code to compute the emission-line luminosities from the \hii{} regions surrounding the star clusters. Mappings solves the equations of thermal equilibrium and ionization balance for a nebula with specified physical conditions, yielding the strengths of key optical and UV emission lines.

In the configuration of the model, we assume a spherical geometry with constant internal pressure throughout the nebula. This choice is physically motivated by the wind-driven bubble model described in Sect. \ref{subsec:wind_drived}, where the pressure is dynamically determined by the mechanical feedback from the central star cluster. The pressure input for each \hii{} region is calculated directly from Eq. \ref{eq:hii_p}. The end condition is radiation bounded ($n_{\rm H~II}/(n_{\rm H~I}+n_{\rm H~II}$) < 5\%), which means low ionizing photons leakage. However, in fact, a significant fraction of ionizing photons can escape the \hii{} region, which can form Diffuse Ionized Gas (DIG), or even escape to the circumgalactic medium (CGM), contributing to the reionization of the universe. This escaping effect will be discussed in Sect. \ref{subsec:dig}.

The elemental abundances used in the photoionization calculations are drawn directly from the ASTRID simulation for the seven tracked species (C, N, O, Ne, Mg, Si, Fe), supplemented with our estimated sulfur abundance as described in Sect. \ref{subsubsec:gas}. Thus, the metallicity of stars and \hii{} regions is derived from star particles and gas particles, respectively. In our photoionization modeling, dust extinction is neglected to simplify our model. This is physically motivated by observations showing that galaxies at $z>6$ are significantly bluer and have a lower infrared excess compared to lower-redshift counterparts \citep[e.g.,][]{Capak2015,Naidu2022,Barger2023}.

The ionizing radiation field is characterized by two key inputs: the $Q$ and SED shape. The shape of the ionizing SED is provided by the interpolated stellar spectra ($L_\lambda(\lambda,t_*,Z_*)$) per $M_{*,0}$ from \texttt{Starburst99}, which naturally account for the dependence on $t_*$ and $Z_*$. The total ionizing photon production rate $Q$ is calculated by integrating $L_\lambda(\lambda,t_*,Z_*)$ from \texttt{Starburst99} SED below 912\AA, and scaled by the small cluster mass $M_{*,0}$ as described in Sect. \ref{subsubsec:star}. 

The starting radius of the photoionization calculation is set to the shell radius $R$ computed from Eq. \ref{eq:hii_r_new}. This self-consistent approach ensures that the ionization parameter $U$ is physically determined by the expansion dynamics of the wind-driven bubble, rather than being treated as an independent free parameter. The ionization parameter is defined as $U = Q/(4\pi R^2 n_{\rm H} c)$, where $n_{\rm H}$ is the hydrogen density at the inner face of the nebula and $c$ is the speed of light. $Q$ depends on the stellar population ($t_*$ and $Z_*$), and the wind-driven bubble model controls $R$ and $n_{\rm H}$ through $L_{\rm mech}$ feedback.

Finally, for each individual star cluster resulting from the splitting procedure described in Sect. \ref{subsubsec:split}, we run \texttt{MAPPINGS V} with the appropriate set of parameters (metallicity, pressure, ionizing SED, and abundances) to obtain the emission-line luminosities. The total emission-line luminosity for each original star particle is then computed by summing contributions from all its constituent small clusters according to Eq. \ref{eq:l_partical}. This approach ensures that we properly account for the statistical distribution of cluster masses while maintaining computational feasibility.

\subsection{Optimizations}

In Sect. \ref{subsubsec:split}, we have already introduced the binning of small cluster's mass distribution which can improve computational efficiency. However, even after filtering by age, there are thousands of young star particles in some massive star-forming galaxies waiting to be computed. Therefore, we adopt an additional optimization strategy: clustering the star particles within each galaxy based on their physical properties, and computing photoionization models only for representative particles from each cluster type.

We employ the \texttt{KMeans} clustering algorithm from the \texttt{sklearn} package. The clustering strategy is tailored to galaxy size: for galaxies with $100 \leq N_{\rm *,<12Myr} \leq 1000$ young star particles, we use $N_{\rm cl}$=16 clusters; for galaxies with $N_{\rm *,<12Myr} > 1000$, we use $N_{\rm cl}$=32 clusters; and for smaller systems with $N_{\rm *,<12Myr} < 100$, we do not perform clustering and compute all particles individually.

The feature vector for clustering includes the following physical parameters: gas-phase elemental abundances (C, N, O, Ne, Mg, Si, Fe), initial gas density ($\rho_0$), star particle mass ($M_*$), stellar age ($t_*$), and stellar metallicity ($Z_*$). Prior to clustering, abundances, density, and mass parameters are all converted to logarithmic scale, and all features are normalized using \texttt{StandardScaler}. Recognizing that stellar age and oxygen abundance have particularly strong influences on emission-line ratios, we assign higher weights to these parameters: $t_*$ receives a weight of 5, log(O/H) a weight of 2, which can produce the lowest relative error in our validation. However, our test shows that the impact of the variation in the weights is very weak on the emission line ratios. If we set all weights to 1, the relative change will be only $\sim$2\%.

This clustering approach reduces thousands of individual photoionization calculations to just 16-32 representative models per galaxy. For each cluster, we compute the mean values of all physical parameters and run a single \texttt{MAPPINGS V} calculation. The total emission-line luminosity for the galaxy is then obtained by summing the contributions from all clusters:

\begin{equation}
L_{\rm gal} = \sum_{i=0}^{N_{\rm cl}-1} n_i L_i, \label{eq:l_gal}
\end{equation}

where $n_i$ is the number of star particles in cluster $i$, and $L_i$ is the emission-line luminosity computed for the representative particle of that cluster, which can be obtained from Eq. \ref{eq:l_partical}.

\begin{figure}
	\includegraphics[width=\columnwidth]{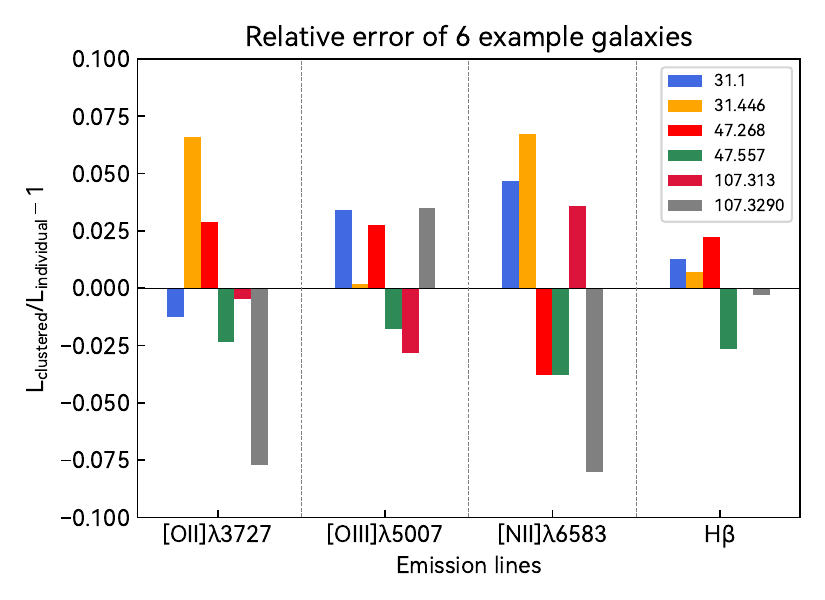}
    \caption{The relative error ($L_{\rm clustered}/L_{\rm individual}$-1) of our clustering optimization. We select 6 example galaxies ([snapshot id].[catalog index]: 31.1, 31.446, 47.268, 47.557, 107.313, and 107.3290), with 2864, 372, 1255, 208, 2447, and 316 young star particles respectively.}
    \label{fig:test_clustering}
\end{figure}

To validate this approach, we select several test clustered galaxies and compare the total emission-line luminosities computed with and without clustering. As shown in Fig. \ref{fig:test_clustering}, the relative deviation is consistently $\lesssim$5\% for all major emission lines, demonstrating that our clustering method maintains accuracy while providing substantial computational savings. This optimization makes our large-scale theoretical calibration project computationally feasible without sacrificing physical fidelity.

\section{Validations of the Modeling Framework}

\subsection{Reproduction of the SFR indicators}

As a first validation of our nebular emission modeling, we compare the simulated star formation rates (SFRs) with commonly used emission-line SFR indicators. These tracers—including UV, IR, H$\alpha$, and \oii{} luminosities—are empirically calibrated to follow relations of the form \citep{Kennicutt1998}:
\begin{equation}
\log{(\rm SFR/(M_\odot~yr^{-1}))} = \log{L_x} - \log{C_x}, \label{eq:sfr_indicator}
\end{equation}
where $x$ denotes the type of tracer, and $C_x$ is the conversion factor between SFR and the relevant luminosity.

We first test the consistency of our model using the \ha{} luminosity. The SFRs of galaxies are derived directly from the ASTRID catalog, while the corresponding H$\alpha$ luminosities are predicted by our photoionization model. Because H$\alpha$ emission arises from the recombination of hydrogen ionized by short-lived O- and B-type stars, it is widely regarded as one of the most reliable tracers of instantaneous star formation. 

\begin{figure*}
	\includegraphics[width=\textwidth]{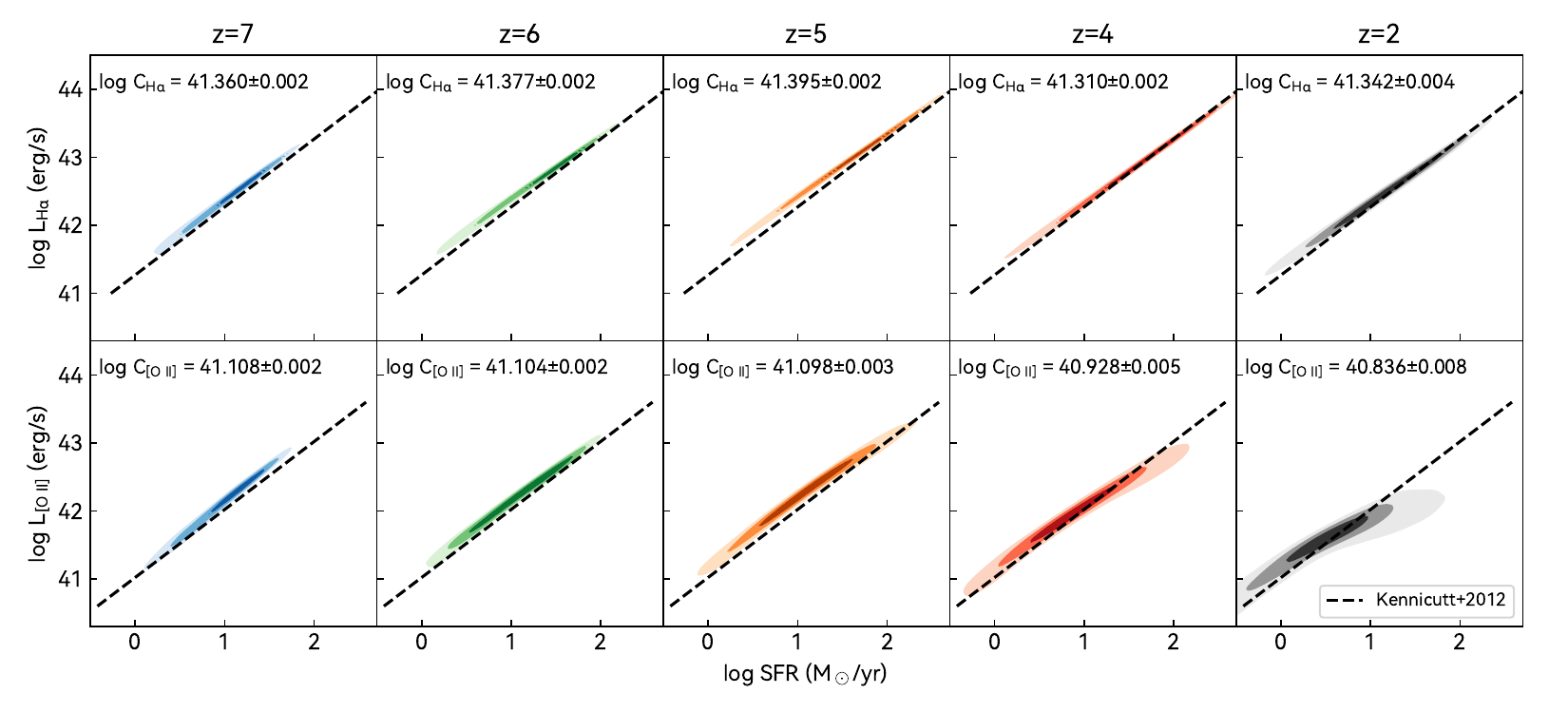}
    \caption{Comparison of SFR from catalog with model-predicted \ha{} (top row) and \oii{} (bottom row) luminosity at redshifts $z$=7,6,5,4,2. The colored contours are our results and the conversion factors ($C_{\rm H\alpha}$ and $C_{\rm [O~II]}$) of SFR are also displayed in all panels. The dashed lines indicate the canonical relationship \citep{Kennicutt1998,Kennicutt2012}.}
    \label{fig:sfr_indicator}
\end{figure*}

For a Kroupa IMF, \cite{Kennicutt2012} gives a canonical conversion factor of $C_{\rm H\alpha}=41.27$. Fitting the relation between the simulated SFRs and the model-predicted H$\alpha$ luminosities yields an average $C_{\rm H\alpha}=41.357$, in excellent agreement with the canonical value. This comparison is shown in the upper five panels of Fig.~\ref{fig:sfr_indicator}.

We perform a similar validation using the \oii{}$\lambda\lambda3726,29$ doublet. Because the ionization potential of oxygen is comparable to that of hydrogen, \oii{} emission can also serve as an empirical SFR tracer under typical ISM conditions \citep{Kennicutt1998}. Adjusting the \cite{Kennicutt1998} calibration to a Kroupa IMF using the H$\alpha$ correction factor from \cite{Kennicutt2012} gives $C_{\rm [O~II]}=41.02$. 

Our model yields a mean conversion factor of $C_{\rm [O~II]}=41.015$, again in excellent agreement with the expected calibration (lower panels of Fig.~\ref{fig:sfr_indicator}).

The close agreement between our model-derived conversion factors and established empirical calibrations provides a strong validation of the nebular emission modeling and supports the reliability of the predicted emission-line luminosities used for subsequent metallicity analysis. 

\subsection{Reproduction of the observed high-redshift luminosity function}

An second validation of our photoionization model is provided by the predicted \oiii$\lambda$5007 luminosity function (LF). While previous work has shown that the ASTRID simulation reproduces the stellar mass function and UV luminosity function of galaxies without any nebular modeling \citep{Bird2022}, here we test whether the model also reproduces the distribution of nebular emission-line luminosities.

\begin{figure}
	\includegraphics[width=\columnwidth]{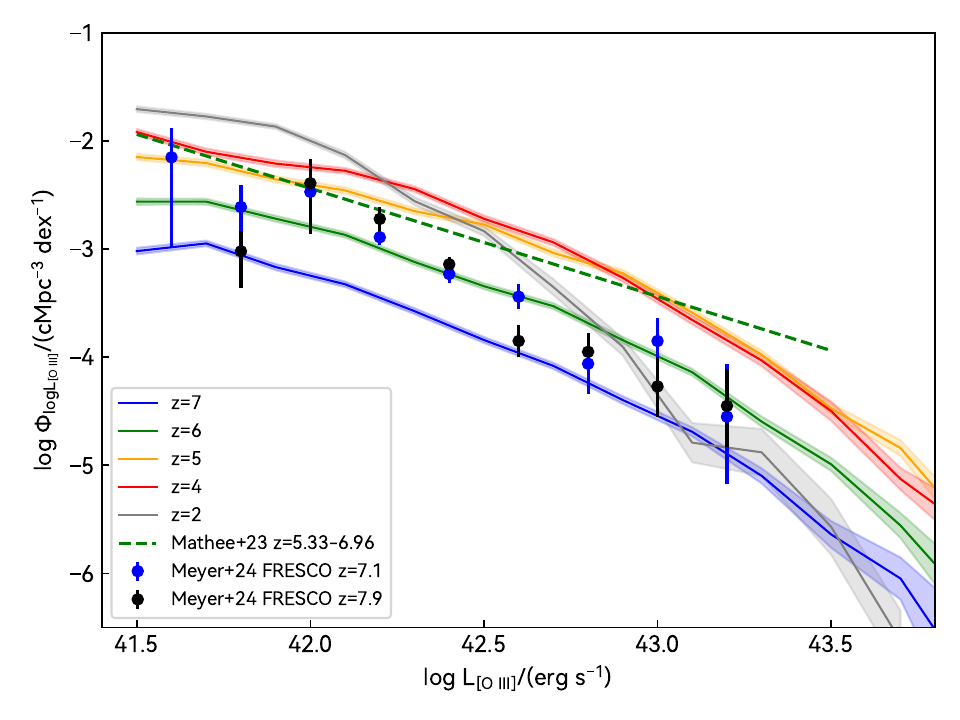}
    \caption{The \oiii$\lambda$5007 luminosity function predicted by our model at redshifts $z$=7,6,5,4,2 (solid lines). The shaded regions represent the combined uncertainty from Poisson counting errors and bootstrap sampling errors. Observational constraints from JWST/NIRCam WFSS at $z$=5.33-6.96 \citep{Matthee2023} are shown as a green dashed line, and the results from the FRESCO survey \citep{Meyer2024} at $z\simeq$7.1 and $z\simeq$7.9 are shown as blue and black dots respectively.} \label{fig:oiii_lum_func}
\end{figure}

Because our galaxy sample is selected uniformly in stellar mass (Sect. \ref{subsubsec:sample}) and is therefore not mass-complete, we apply a completeness correction when estimating the number density $\Phi_{\log{L_{\rm [O~III]}}}$. This correction is obtained by interpolating the ratio of selected galaxies to the total number of galaxies in the full simulation catalog within each $\log{M_*}$ bin.

Fig. \ref{fig:oiii_lum_func} shows the predicted \oiii{} luminosity functions across redshifts, together with recent high-redshift observational constraints. The uncertainties (shaded regions) are derived by combining Poisson errors (based on galaxy counts in each luminosity bin) with statistical sampling errors estimated from 10,000 bootstrap resamples of our galaxy sample.

Overall, the predicted luminosity functions show good agreement in both slope and number density with the available JWST/NIRCam WFSS measurements \citep{Matthee2023} and the FRESCO survey detections \citep{Meyer2024}. This consistency supports the physical realism of our nebular emission line modelling and indicates that the simulation produces galaxy populations with emission-line properties comparable to those observed at high redshift.

\subsection{Reproduction of the observed emission line ratios}

\begin{figure*}
	\includegraphics[width=\textwidth]{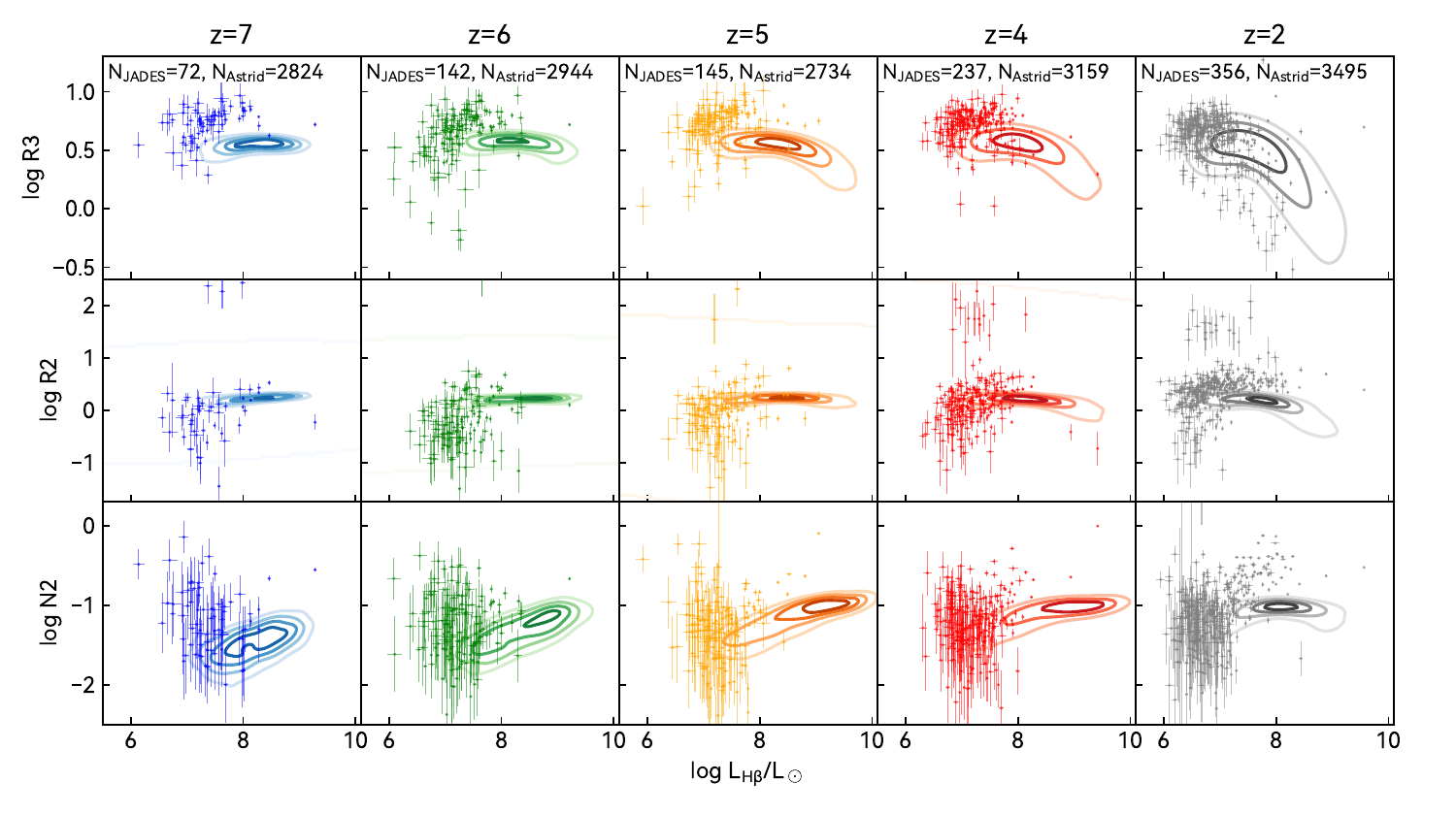}
    \caption{The line ratios (R3=\oiii$\lambda$5007/\hb, R2=\oii$\lambda$3727/\hb, and N2=\nii$\lambda$6583/\ha) at different \hb{} luminosity. The contours are the ASTRID galaxies, whose line ratios are constructed by our model. The dots with error bar are the JWST JADES DR4 sample. The number of JADES galaxies is shown as $N_{\rm JADES}$ in each column. The emission line fluxes of the JADES sample have been corrected for dust extinction.}
    \label{fig:line_ratio}
\end{figure*}

In Fig. \ref{fig:line_ratio}, we present the three most common line ratios (R3=\oiii$\lambda$5007/\hb, R2=\oii$\lambda$3727/\hb, and N2=\nii$\lambda$6583/\ha) at different \hb{} luminosities ($L_{\rm H\beta}$). For comparison, we also present the JWST JADES DR4 sample \citep{Eisenstein2026,CurtisLake2026,Scholtz2026} on the line ratio--$L_{\rm H\beta}$ plane. We use the emission line fluxes from the JADES NIRSpec R1000\_5pix catalog. We only select galaxies with high S/N of \ha{} and \hb{}. For $z\ge5$ galaxies, the S/N cutoff is 3 and for $z\le4$ galaxies, the S/N cutoff is 5. We perform a dust extinction correction using the Balmer decrement with the assumption of the intrinsic \ha{}/\hb{}=2.86 under the case-B recombination. All the fluxes of emission lines with \ha{}/\hb{}>2.86 are corrected with the \cite{Calzetti2000} attenuation curve. For the fluxes with \ha{}/\hb{}$\le$2.86, no correction is performed.

We use the \hb{} luminosity as a practical proxy because stellar masses are not yet available. It may not be an exact substitute for matching by stellar mass. From Fig. \ref{fig:line_ratio}, We can see that our results of R3, R2, and N2 agree well with the JADES sample. Due to the limited comoving volume of JADES, the number of ASTRID galaxies at the high-luminosity end is much more than that of JADES. However, there are systematic biases in certain redshift and luminosity ranges. Specifically, at the high-redshift ($z\ge4$), low-luminosity (or low-mass) end, the R3 and N2 given by our model are systematically lower than those of the JADES samples. The reasons for this will be explained in detail in the discussion section (Sect. \ref{subsec:low_r3} and \ref{subsec:n_abundance}).

\section{Strong line metallicity calibrations}
Having established that the model reproduces both empirical SFR indicators and the observed \oiii{} luminosity function, we now use the simulated galaxy sample to derive theoretical calibrations between strong emission-line ratios and gas-phase metallicity.

\subsection{Optical calibrations} \label{subsec:optical_calibrations}

\begin{figure*}
	\includegraphics[width=\textwidth]{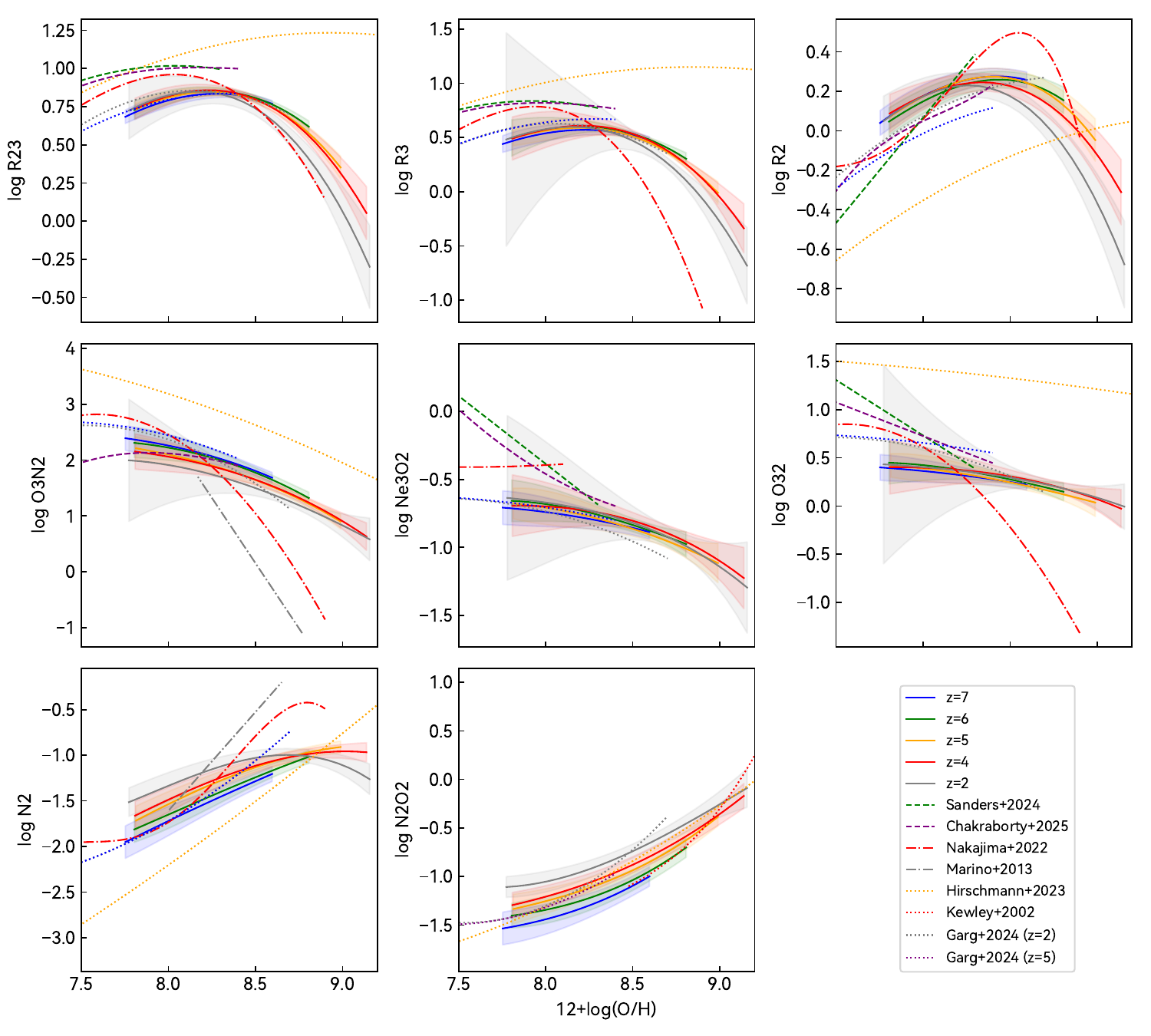}
    \caption{Optical diagnostics of R23, R3, R2, O3N2, Ne3O2, O32, N2 and N2O2. Solid curves is our fitting and the filled regions are the standard deviation of the logarithm of the line ratio at the corresponding metallicity. Dashed, dot-dashed and dotted curves represent high-redshift, low-redshift, and model-based calibrations respectively.}
    \label{fig:optical}
\end{figure*}

In this section, We now examine the relationship between strong optical emission-line ratios and gas-phase metallicity predicted by the ASTRID simulation combined with our photoionization modeling. We focus on three categories of diagnostics, defined as the following equations, the first group is oxygen-based diagnostics (R23, R3, R2):
\begin{itemize}
    \item R23$=\frac{\rm \oiii\lambda\lambda4959,5007+\oii\lambda3727}{\rm \hb}$,
    \item R3$=\frac{\rm \oiii\lambda5007}{\rm \hb}$,
    \item R2$=\frac{\rm \oii\lambda3727}{\rm \hb}$.
\end{itemize}

The second group is nitrogen-base diagnostics (N2, O3N2, N2O2):
\begin{itemize}
    \item N2$=\frac{\rm \nii\lambda6583}{\rm \ha}$,
    \item O3N2$=\frac{\rm \oiii\lambda5007/\hb}{\nii\lambda6583/\ha}$,
    \item N2O2$=\frac{\rm \nii\lambda6583}{\rm \oii\lambda3727}$.
\end{itemize}
    
The last group is ionization-sensitive diagnostics that depend strongly on the ionization parameter (O32, Ne3O2)
\begin{itemize}
    \item O32$=\frac{\rm \oiii\lambda5007}{\rm \oii\lambda3727}$,
    \item Ne3O2$=\frac{\rm \neiii\lambda3869}{\rm \oii\lambda3727}$.
\end{itemize}

To quantify the theoretical calibration relations, we perform a cubic polynomial fit in the form of $y=a_0+a_1x+a_2x^2+a_3x^3$, where $x$ is 12+log(O/H) and $y$ is the logarithm of the line ratio, using an ordinary least squares regression. Fig. \ref{fig:optical} shows the fitting results at different redshifts (solid curves). The intrinsic scatter of the models---calculated as the standard deviation of the line ratio within bins of 12+log(O/H) and smoothed with a cubic polynomial--is represented by the shaded regions. For context, we compare our theoretical predictions with existing model-based calibrations \citep[dotted curves;][]{Kewley2002,Hirschmann2023,Garg2024} and direct $T_e$-based measurements from both low-redshift \citep[dash-dotted curves;][]{Marino2013,Nakajima2022} and high-redshift \citep[dashed curves;][]{Sanders2024,Chakraborty2025} observations. The best-fit coefficients are tabulated in Appendix \ref{tab:optical_table}.

\subsubsection{Common diagnostics--R23, R3, and R2}
The top row of Fig. \ref{fig:optical} confirms the characteristic double-branched nature of the R23, R3, and R2 diagnostics. The turnover point for R23 occurs at 12+log(O/H)$\simeq$ 8.3 and log R23$\simeq$0.8. The movement of the turnover point of R23 is weak ($z=7$: 8.275, $z=6$: 8.250, $z=5$: 8.238, $z=4$: 8.233, $z=2$: 8.168). Notably, the turnover metallicities for R3 and R2 are offset to lower and higher values, respectively, consistent with their dependence on different ionization states of oxygen.

Our predictions for R23 and R3 align closely with those from \cite{Garg2024} but lie systematically below the recent high-redshift $T_e$-based calibrations of \cite{Sanders2024} and \cite{Chakraborty2025}, as well as the model of \cite{Hirschmann2023}. Conversely, our R2 ratios are systematically higher than those works. The most pronounced deviation is from \cite{Hirschmann2023}. This pattern--elevated R2 and suppressed R3--points to systematically lower ionization parameters ($U$) in our models compared to theirs, a conclusion further supported by the O32 diagnostic (see below). This discrepancy likely stems from fundamental differences in determining $U$: our wind-driven bubble model dynamically computes $U$ from stellar feedback and local density, whereas \cite{Hirschmann2023} adopt different stellar ionizing spectra (e.g., via a higher upper mass limit of 300 M$_\odot$ for the IMF) and empirical $U$-Z relations from \cite{Carton2017}. A more detailed discussion about the systematically lower R3 is in Sect. \ref{subsec:low_r3}.

A key finding is the superior stability of the R23 diagnostic. The scatter (shaded regions) for both R2 and R3 is substantially larger than for R23, particularly at the low-metallicity end of R3 at $z=2$. Furthermore, the redshift evolution of R2 and R3 is more pronounced compared to R23. This is expected because R23 combines two major ionization states of oxygen (\oii{} and \oiii{}), making it less sensitive to variations in $U$ than ratios involving only a single state.

\subsubsection{Nitrogen-based diagnostics--N2, O3N2, N2O2} \label{subsubsec:n_calibration}

Nitrogen-based diagnostics are widely used to determine metallicity of local galaxies \citep[e.g.,][]{Pettini2004,PerezMontero2009,Marino2013}. Their popularity stems from the nearly monotonic relationship between nitrogen line ratios and metallicity, as well as the observational constrains of ratios such as N2, which involves lines separated by only a small wavelength interval and therefore requires minimal dust correction and are almost unaffected by $U$ due to the closed first ionization energy of H, N, and O.

The three nitrogen-based diagnostics shown in the lower-left panels of Fig. \ref{fig:optical} exhibit stronger redshift evolution than the oxygen-based R23 ratios discussed above. Compared with other results, at the low-metallicity end, our predictions agree reasonably well with existing calibrations. However, at higher metallicities, our predicted \nii{} luminosity are systematically weaker, resulting in a flatter slope for all N-based diagnostics. The notable exception is the model of \cite{Hirschmann2023}, from which we deviate across the entire metallicity range. It is the same as the systematic deviation found in R3 and R2, caused by the IMF and $U$.

This slope discrepancy arises from the nitrogen-to-oxygen abundance ratio predicted by the ASTRID simulation, which does not fully reproduce realistic chemical evolution. As a result, the nitrogen-based calibrations should be interpreted with caution. We explore the implications of this issue in more detail in Sect. \ref{subsec:n_abundance}.


\subsubsection{Ionization-sensitive diagnostics--O32, Ne3O2} \label{subsubsec:optical_o32}

Diagnostics like O32 and Ne3O2, which primarily trace the ionization parameter $U$, exhibit larger scatter and systematic offsets when used as indirect metallicity indicators (mid-right panels of Fig. \ref{fig:optical}). We find that these indicators, which are positively correlated with ionization parameter ($U$), are negatively correlated with gas-phase metallicity ($Z_g$). This aligns with the theoretical expectation of a correlation between $U$ and metallicity \citep[$U \propto Z^{-0.8}$;][]{Dopita2006,Carton2017}. However, it is important to note that the dependence of O32 on $Z_g$ is quite complex and depends on the specific \hii{} region model. O32 depends not only on $U$ but also on $T_e$ and $n_{\rm H}$. $T_e$ depends on $Z_g$, while $U$ in our model depends on $t_*$, $Z_*$, and $n_{\rm H,0}$. This is why ionization-sensitive calibrations exhibit such large dispersion.

The O32 at the high-metallicity end of \cite{Nakajima2022}'s curve (red dot-dashed curve in Fig. \ref{fig:optical}) is quiet lower than all of other high-redshift curves including ours. The data of \cite{Nakajima2022} at the high-metallicity end is from the local galaxies in Sloan Digital Sky Survey DR7 calibrated by the direct-$T_e$ method \citep{Curti2020}. This implies the complex dependence of O32 on the environment; even with the same $Z_g$, O32 can vary drastically under different environments at different redshift.

The behavior of O32 and Ne3O2 corroborates the systematic differences found in the R2 and R3 diagnostics. Specifically, a combination of lower-than-expected R2 and higher-than-expected R3 corresponds to higher predicted O32 and Ne3O2 ratios. Since O3N2 = log(R3/N2), it can also be affected by these $U$-driven offsets, explaining its underestimation relative to \cite{Hirschmann2023}. The model dependence, substantial scatter, and systematic uncertainties in these $U$-sensitive ratios highlight the challenge of using them as precise metallicity diagnostics.

\subsection{UV calibrations}

Rest-frame UV spectroscopy has become an increasingly powerful tool for probing the ISM of galaxies at $z\gtrsim4$, where traditional optical diagnostics shift into difficult-to-access infrared regimes. High-redshift galaxies typically exhibit hard radiation fields powering prominent high-ionization lines (e.g., \ciiisemi$\lambda\lambda$1907,09 and \oiiisemi$\lambda\lambda$1661,66) that serve as tracers of density, ionization, and chemical abundances \citep[e.g.,][]{Erb2010,Berg2016,Feltre2016,PerezMontero2017,Acharyya2019,Kewley2019,Mingozzi2022}. In particular, utilizing these UV features to derive the carbon and oxygen abundance can provide critical constraints on galaxy chemical evolution in the early universe, offering a vital alternative to optical methods.

\begin{figure*}
	\includegraphics[width=\textwidth]{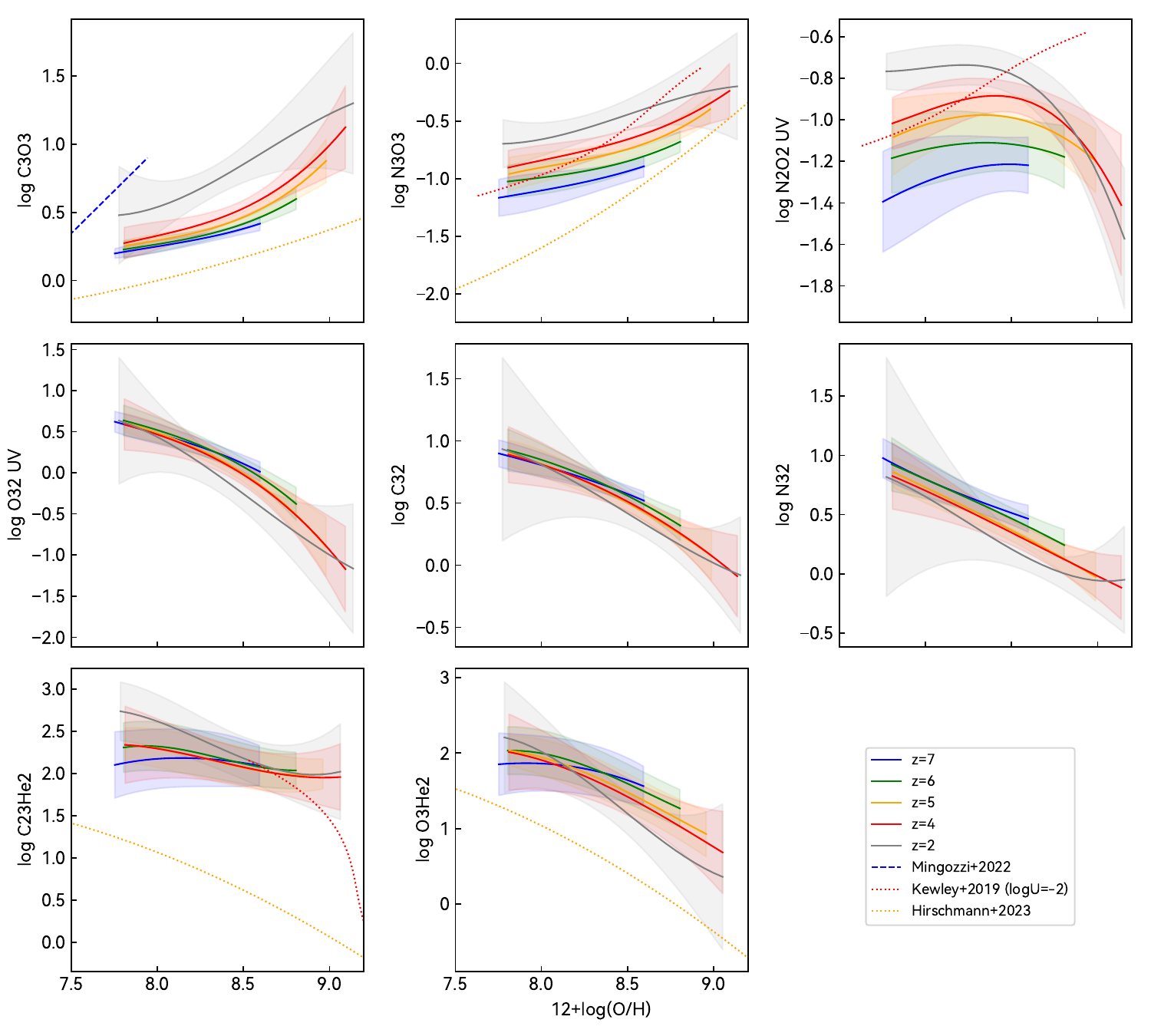}
    \caption{UV diagnostics of C3O3, N3O3, N2O2 UV, O32 UV, C32, N32, C23He3, and O3He2. The legend is the same as Fig. \ref{fig:optical}.}
    \label{fig:uv}
\end{figure*}

Here we use the ASTRID simulation to derive theoretical calibrations between several commonly used UV emission-line ratios and gas-phase metallicity. These relations provide a complementary framework to optical diagnostics for interpreting the rapidly growing sample of high-redshift galaxies observed with JWST. 
In this section, we present the common rest-frame UV diagnostics of metallicity derived from the ASTRID simulation and our photoionization model including C3O3, N3O3, N2O2 UV, O32 UV, C32, N32, C23He2, and O3He2, defined as the following equations:
\begin{itemize}
    \item C3O3$=\frac{\rm \ciiisemi\lambda\lambda1907,09}{\rm \oiiisemi\lambda\lambda1661,66}$,
    \item N3O3$=\frac{\rm \niii\lambda1750}{\rm \oiiisemi\lambda\lambda1661,66}$,
    \item N2O2 UV$=\frac{\rm \nii\lambda\lambda2139,43}{\rm \oii\lambda2470}$,
    \item O32 UV$=\frac{\oiiisemi\lambda1666}{\oii\lambda2470}$,
    \item C32$=\frac{\rm \ciiisemi\lambda\lambda1907,09}{\cii\lambda2325}$,
    \item N32$=\frac{\rm \niii\lambda1750}{\rm \nii\lambda\lambda2139,43}$,
    \item C23He2$=\frac{\rm \ciiisemi\lambda\lambda1907,09+\cii\lambda2325}{\rm \heii\lambda1640}$,
    \item O3He2$=\frac{\rm \oiiisemi\lambda\lambda1661,66}{\rm \heii\lambda1640}$.
\end{itemize}

We perform the same fit as the optical calibrations for the UV, shown in Fig. \ref{fig:uv}, and the coefficients are shown in the Table \ref{tab:uv_table} in the Appendix. Due to the relatively limited exploration of UV diagnostics in the literature, we only compare our results primarily with the photoionization model grids of \cite{Kewley2019}, the direct $T_e$-based measurements of \cite{Mingozzi2022}, and the simulation-based results of \cite{Hirschmann2023}.

Both C3O3 and N3O3 show monotonic relationships with metallicity, though with considerable intrinsic scatter (Fig. \ref{fig:uv}, top two diagrams). Notably, the systematic offset from \citet{Hirschmann2023} persists in the UV, similar to the trends seen in optical diagnostics, but N3O3 follows the high ionization parameters model ($\log{U}=-2$) of \cite{Kewley2019} well. For ionization-sensitive line ratios such as O32 UV, C32, and N32, they also show a clear dependence on metallicity and minor dependence on redshift, but the dispersion is large, especially at $z$=2, just like the optical case of O32 and Ne3O2 (Sect. \ref{subsubsec:optical_o32}).

The diagnostics involving \heii$\lambda$1640--C23He2 and O3He2--present a more complex picture (Fig. \ref{fig:uv}, bottom twe diagrams). While a general negative trend with metallicity is discernible, both ratios exhibit extremely large scatter, rendering them unreliable as precise metallicity indicators in our framework. It is also important to notice that we only considered \hii{} regions. Actually, the \heii$\lambda$1640 line can be powered by multiple, highly variable sources in high-redshift galaxies: hard radiation from Wolf-Rayet stars, residual accretion onto massive stars, or even a contribution from active galactic nuclei (AGN). The above process may introduce much more scatter. Therefore, we do not recommend using these diagnostics.

A significant and unexpected finding concerns the N2O2 UV diagnostic. While \citet{Kewley2019} identifies N2O2 UV as one of the most promising UV metallicity tracers, our models predict it to be largely insensitive to metallicity, exhibiting the largest scatter and strongest redshift dependence among the diagnostics studied here (Fig. \ref{fig:uv}, top-right diagram). We caution that this result is likely non-physical and directly tied to the problematic N/O abundance ratio in the ASTRID simulation, as shown in Sect. \ref{subsubsec:n_calibration} and further analysed in Sect. \ref{subsec:n_abundance}. Therefore, our N2O2 UV relation should not be used as a reliable calibration until the nitrogen enrichment in simulations is better understood and calibrated.

\section{Discussion}

\subsection{Comparison with pre-computed grids}

To illustrate the advantages of our approach, we compare our results with those of pre-computed grids. We use the \texttt{Synthesizer} code \citep{Lovell2025,Roper2026} for comparison. \texttt{Synthesizer} is an open-source python package for generating synthetic observations from theoretical models. It includes some built-in photoionization grids calculated by \texttt{Cloudy}. Users can input data of star particles from cosmological simulations into \texttt{Synthesizer} and then obtain the emission line luminosity of galaxies.

\begin{table*}
	\centering
	\caption{Comparison of theoretical framework}
	\label{tab:method_compare}
	\begin{tabular}{c|C{2.9cm}C{2.9cm}C{4.1cm}C{2.9cm}} 
		\hline
		  Parameter & This research & Synthesizer & \cite{Hirschmann2023} & \cite{Garg2024} \\
		\hline
        Simulation & ASTRID & ``Arbitrary'' & IllustrisTNG & SIMBA \\
        \hline
        Redshift & $z$=2-7 & ``Arbitrary'' & $z$=0-8 & $z$=1-5 \\
        \hline
        Photoionization code & MAPPINGS V 5.2.0 & grid pre-computed by Cloudy 23.01 & grid pre-computed by Cloudy 13.03 \citep{Gutkin2016} and MAPPINGS V (fast radiative shock) & Cloudy 17.00 \\
        \hline
        SSP synthesis & Starburst99 & BPASS 2.2.1 & BC03 & BPASS 2.1 \\
        IMF & Kroupa (max 100 M$_\odot$) & Chabrier (max 300 M$_\odot$) & Chabrier (max 300 M$_\odot$) & Chabrier (max 100 M$_\odot$) \\
        \hline
        Sub-grid CMF & $\beta$=-2, $M_\mathrm{min}$=2$\times10^3$, $M_\mathrm{max}$=4$\times10^5$ & No & No & $\beta$=-2, $M_\mathrm{min}$=10$^{3.5}$, $M_\mathrm{max}$=10$^{5.0}$ \\
        \hline
        Gas density ($n_\mathrm{H}$) & regulated by $\rho_0$ and stellar feedback & 1000 cm$^{-3}$ & 100 cm$^{-3}$ & 30 cm$^{-3}$ \\
        \hline
        Gas metallicity ($Z_g$) & from gas particles & $Z_g=Z_*$ & from gas particles & $Z_g=Z_*$ \\
        \hline
        N/O & from gas particles & \cite{Nicholls2017} & determined by Gutkin's grid & \cite{Pilyugin2012} \\
        \hline
        Ionization parameter (U) & regulated by model & parameterized by $t_*$ and $Z_*$ \citep{Wilkins2020} & parameterized by SFR and filling factor \citep{Hirschmann2017} & regulated by sub-grid \\
        \hline
        Dust & No & No & Yes & Yes \\
        DIG & Yes & No & No & Yes \\
        Post-AGB & No & No & Yes & Yes \\
        AGN & No & No & Yes & No \\
        
		\hline
	\end{tabular}
\end{table*}

We use the default grid configuration of \texttt{Synthesizer} for testing. In the default grid, the gas density is fixed at 1000 cm$^{-3}$, and the ionization parameter ($U$) uses the empirical relationship from \cite{Wilkins2020}, controlled by the $t_*$ and $Z_*$ of star particles, with a reference value of $\log{U_{\rm ref}}=-2$, anchored at $t_*=1$ Myr and $Z_*=0.01$. Because Synthesizer only collects information from stars and uses a fixed gas density and empirical ionization parameter relationship. We find that the metallicity calibrations derived from the default grid lack redshift evolution. Therefore, it is not suitable for metallicity diagnostic in high-redshift galaxies. To address this issue, it might be necessary to use multiple different grids to cover higher-dimensional parameter spaces to match ionization environments at different redshifts; however, this would lead to excessive computation and inefficient sampling. For comparison, we show the parameters of our method, the \texttt{Synthesizer}, and two other simulation-based researches \citep{Hirschmann2023,Garg2024} in Table \ref{tab:method_compare}.

\subsection{The AGN contamination}

Recently, JWST has discovered many little red dots (LRDs) at high redshift ($z$>4) \citep[e.g.,][]{Labbe2023,Kocevski2023,Kokorev2023,Barro2024,Matthee2024}. Spectroscopic observations reveal that most LRDs exhibit broad lines \citep[e.g.,][]{Greene2024,Hviding2025}, which could be evidence of active galactic nuclei (AGN). Because photoionization modeling of AGN is complex and the ionizing photon budget during the reionization epoch is dominated by star formation rather than AGN \citep[e.g.,][]{Robertson2015,Dayal2025,Jiang2025}, the detailed emission-line properties of AGN are not the focus of this paper. Therefore, instead of modeling the AGN emission independently, we adopt an empirical relationship between the bolometric luminosity ($L_{\rm bol}$) of the black hole and the \oiii{} emission-line luminosity ($L_{\rm[O~III]}$) from \cite{Heckman2004} in the form of $L_{\rm bol}=3500L_{\rm[O~III]}$ to estimate the \oiii{} luminosity contributed by the AGN.

$L_{\rm bol}$ can be determined by the accretion rate of central black hole ($\dot{M}$):
\begin{equation}
L_{\rm bol} = \eta\dot{M}c^2, \label{eq:l_bol}
\end{equation}
where $\dot{M}$ is from the ASTRID black hole particle catalog and the radiative efficiency $\eta=0.1$. We choose the nearest black hole particle to the galaxy center as the central super-mass black hole.

We find that the AGN activity in our sample is quite weak, especially at the highest redshift. The fraction of galaxies with an \oiii{} luminosity contribution from AGN greater than 10\% at $z$=7,6,5,4,2 are 0\%, 0.2\%, 0.7\%, 4.5\%, 15.6\% respectively. Therefore, AGN is expected to have a negligible impact on our luminosity function and metallicity calibrations. However, we need to point out that although the bolometric conversion to \oiii{} from \cite{Heckman2004} is locally calibrated and may not hold at $z$>4 due to different covering factors and obscuration, this is not enough to change our conclusion on weak contamination of AGN.


\subsection{The DIG effect} \label{subsec:dig}

Diffuse Ionized Gas (DIG) significantly contributes to the total line emission in galaxies, typically exhibiting lower ionization states than compact \hii{} regions \citep{Haffner2009}. While some frameworks model the DIG as a distinct physical component with a fixed escape fraction \citep[e.g.,][]{Garg2024} ($f_{\rm esc}$), our approach incorporates it self-consistently in the \cite{Dopita2006} wind-driven model.

In this framework, the DIG is the natural evolutionary end-state of an \hii{} region. As a stellar cluster ages ($> 3$--$5$ Myr), the ionizing flux wanes while the region expands due to mechanical feedback. This leads to a rapid decline in the ionization parameter, physically manifesting as the low-surface-brightness, low-ionization gas characteristic of the DIG.

By integrating emission over the full \hii{} region lifecycle--from early high-pressure phases to late diffuse stages--the DIG is inherently captured in our synthetic spectra. This eliminates the need for independent DIG modeling. However, our model does not yet account for the $\sim$10\% of ionizing photons that may leak through low-density ISM channels into the CGM \citep{Rosdahl2018, Steidel2018}, which may further power extended emission or contribute to cosmic reionization.

\subsection{The underestimation of R3 at the low-mass end} \label{subsec:low_r3}

We find that our model exhibits a systematic underestimation of the R3=\oiii$\lambda$5007/\hb{} ratio at the low-mass end (see Figure \ref{fig:line_ratio} and \ref{fig:optical}). Specifically, our predicted values cluster around log R3 $\sim 0.5-0.6$, whereas recent JWST observations from the JADES survey reveal a more highly ionized ISM with log R3 $\sim 0.7-0.8$ \citep[e.g.,][]{Cameron2023a,Sanders2023,Scholtz2026}. Rather than being driven by a single isolated factor, this mild tension most likely reflects a synergetic combination of multiple physical mechanisms and sub-grid limitations at cosmic dawn:

First, the discrepancy may stem from variations in the effective cluster mass distribution and spatial clustering. In our baseline model, we adopt a standard CMF power-law index of $\beta = -2$, and star particles are treated as isolated entities driving independent wind bubbles. The ionization parameter will increase with the cluster mass ($U\propto M_{*,0}^{1/5}$). However, in the intense, high-pressure environments of early galaxies, the CMF might be flatter (i.e., $\beta>-2$), shifting the cluster population toward higher-mass clusters \citep{Adamo2020}. Furthermore, because $z > 6$ star-forming galaxies feature exceptionally high star formation rate surface densities ($\Sigma_{\rm SFR}$), the spatial proximity of multiple young star clusters in extremely compact environments inevitably leads to the merging of individual \hii{} regions into giant, co-spatial "super-bubbles". This nebular overlap mechanism effectively boosts the collective ionizing photon output per unit volume, acting in the same direction as a flatter CMF by increasing the effective cluster mass. As presented in our sensitivity tests in Appendix \ref{apdx:split}, while systematically varying these CMF-related sub-grid parameters within our current framework shows a relatively minor isolated impact on the integrated line ratios, their real-world physical coupling under extreme high-redshift conditions could significantly elevate the local ionization parameter ($\log U$) and consequently boost the R3 ratio.

Second, the limitation of our adopted SPS templates could contribute to the softer radiation field. Currently, our framework relies on \texttt{Starburst99} templates governed by standard Pop II stellar evolutionary tracks and a conventional Kroupa/Chabrier IMF. At cosmic dawn ($z > 6$), however, the contribution from pristine Pop III stars ($Z_*=0$) or a top-heavy IMF dominated by extremely massive stars ($>100\,M_\odot$) might still be significant \citep[e.g.,][]{Vanzella2020,Maiolino2024,Cullen2025,Fujimoto2025}. A top-heavy IMF or a Pop III component provides a much harder ionizing spectrum, substantially increasing the production of high-energy photons ($>35.12\rm\,eV$) required to ionize ${\rm O}^+$ to ${\rm O}^{2+}$. Incorporating more flexible SPS models like BPASS or specific Pop III modules could naturally resolve the suppressed R3 issue.

Third, our photoionization modeling assumes a strict radiation-bounded geometry with zero ionizing photon escape ($f_{\rm esc} = 0$). Galaxies at $z > 6$ are the primary agents of cosmic reionization, which strongly implies non-zero and potentially high escape fractions. If a significant fraction of \hii{} regions are density-bounded, the outer, low-ionization layers of the nebulae—where \oii{} emission and a substantial portion of Balmer lines originate—are truncated due to photon leakage \citep[e.g.,][]{Jaskot2013,Wang2019,Ding2023}. Because \oiii{} is concentrated closer to the central ionizing source, a density-bounded structure preferentially suppresses the integrated \hb{} and \oii{} fluxes, shifting the integrated spectrum toward higher R3 ratios. Since our absolute luminosity of \oiii{} is in good agreement with the observations, the reduction in the luminosity of low-ionized lines due to ionizing photon escape may be the main contributor of R3.

In summary, the underpredicted R3 in low-mass galaxies at $z>6$ points toward an intriguing transition in the physical state of the ISM at cosmic dawn. The true observed high-excitation states are likely the cumulative result of dense nebular merging under a flatter cluster mass distribution, harder stellar radiation fields, and porous geometries acting simultaneously. Future updates incorporating high-resolution radiation-hydrodynamic simulations that resolve multi-phase ISM structures, coupled with Pop III-inclusive stellar templates, will be essential to fully alleviate these discrepancies.

\subsection{The nitrogen abundance} \label{subsec:n_abundance}

\begin{figure*}
	\includegraphics[width=\textwidth]{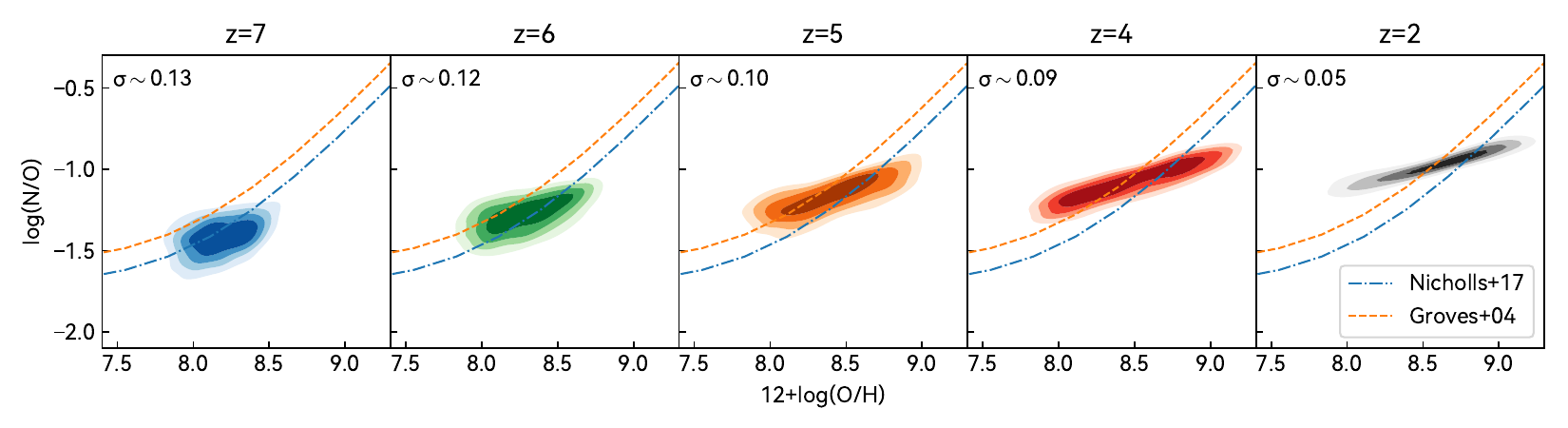}
    \caption{The N/O-O/H relationship of our ASTRID sample. Dashed  \citep{Groves2004} and dot-dashed \citep{Nicholls2017} curves represent the relationship in local universe. $\sigma$ is the average standard deviation of log(N/O) at given O abundance.}
    \label{fig:nitrogen}
\end{figure*}

A significant finding in our analysis is that nitrogen-based diagnostics (e.g., N2, O3N2, N2O2) exhibit a systematically flatter slope and weaker line intensities at high metallicities compared to other theoretical and observational calibrations (see Sect. \ref{subsec:optical_calibrations}). This discrepancy points to a limitation in the nitrogen enrichment modeling in the ASTRID simulation. Fig. \ref{fig:nitrogen} illustrates the N/O-O/H relationship of our sample. At $z=2$, the N/O ratio at high oxygen abundances is clearly lower than the relationships observed in the local universe, which results to a negligible increase in \nii{} luminosity with increasing O/H and causes a non-physical turnover in the N2 metallicity diagnostic.

The accuracy of nitrogen and carbon abundances in cosmological simulations is primarily limited by the uncertainties in stellar yield tables and the numerical treatment of the enrichment process. In ASTRID, the metal return from massive stars ($8-40 M_{\odot}$) and AGB stars ($1-8 M_{\odot}$) follows \cite{Kobayashi2006} and \cite{Karakas2010}, respectively \citep{Bird2022}. However, yields for nitrogen and carbon in low-metallicity AGB stars are notoriously uncertain, with model-dependent variations reaching up to orders of magnitude \citep{Karakas2014}. Furthermore, ASTRID employs a linear scaling for the $8-13 M_{\odot}$ mass range due to the lack of tabulated data in the original \cite{Kobayashi2006}'s tables \citep{Bird2022}. This mass regime corresponds to Super-AGB stars, which are critical contributors to primary nitrogen production via hot-bottom burning, and a simplified linear treatment may fail to capture the non-linear enrichment required to match high-redshift observations.

Moreover, recent JWST/NIRSpec observations have revealed a systematic nitrogen enhancement in metal-poor galaxies at high redshift \citep[e.g.,][]{Bunker2023,Cameron2023,Ji2024,Schaerer2024,Cataldi2026}. The reasons for this N-enrichment phenomenon are still being studied and some studies question its existence \citep[e.g.,][]{Zhu2025,Schaerer2026}. \cite{Cataldi2026} uses direct $T_e$-based abundance and finds that compared to local relations, galaxies at $z$=1-6 exhibit a median N/O offset of $\sim 0.18$ dex at fixed O/H, which increases to $0.3-0.4$ dex in metal-poor environments (12+log(O/H)$\le$ 8.1). From the N/O-O/H relationship in Fig. \ref{fig:nitrogen} we can see that ASTRID's chemical network does not reproduce this ``N-enhanced'' population, leading to a systematic bias in our derived calibrations. High-resolution zoom simulations like THESAN-ZOOM \citep{McClymont2026} suggest that bursty star formation and temporally differential galactic winds can trigger order-of-magnitude excursions in N/O on timescales of $\le 100$ Myr. Since ASTRID's SPH-based hydrodynamics lack a particle-splitting mechanism to facilitate metal diffusion, the simulation likely smooths out these extreme local enrichment events. We use the result of \cite{Cataldi2026} to estimate the systematic bias if we apply our N2 and N2O2 calibrations to a real high-z JWST galaxy. We find that our N2 and N2O2 calibrations will overestimate the oxygen abundance by $\sim$0.31 and $\sim$0.35 dex at $z=4$, respectively. Consequently, the nitrogen and carbon line predictions in this work should be treated with caution, as they may not fully represent the chemical diversity of the high-redshift interstellar medium, although the additional enhancement of carbon is not evident at high redshift.

Another related concern is whether the nitrogen abundance could affect the nebular cooling efficiency. At high metallicities (12+log(O/H)>8.5), cooling through optical and infrared nitrogen lines can become non-negligible, which could in principle introduce a bias in the derived $T_e$ and emission line ratios. To quantify this effect, we performed a controlled test by artificially adopting the N/O–O/H relation from \cite{Nicholls2017} in place of the ASTRID nitrogen abundance for a subsample of galaxies at $z=2$ (where the N abundance is the highest). The resulting $T_e$ only differs $\sim$50 K on average from the default models, well below the typical measurement uncertainty of $\sim$500-1000 K in high-redshift direct-$T_e$ studies. We therefore conclude that the impact of nitrogen abundance uncertainties on the nebular thermal balance is negligible.

\subsection{The best choice of optical calibration}

In our results, we find that the dispersions of optical calibrations are significantly better than that of UV calibrations. Therefore, we recommend using optical emission lines to calibrate the metallicity. However, the optimal choice among specific methods requires evaluation, which is discussed in this section.

\subsubsection{Bayesian framework and error estimation}

\begin{figure}
	\includegraphics[width=\columnwidth, trim={1.3cm 0cm 1.3cm 0cm}, clip]{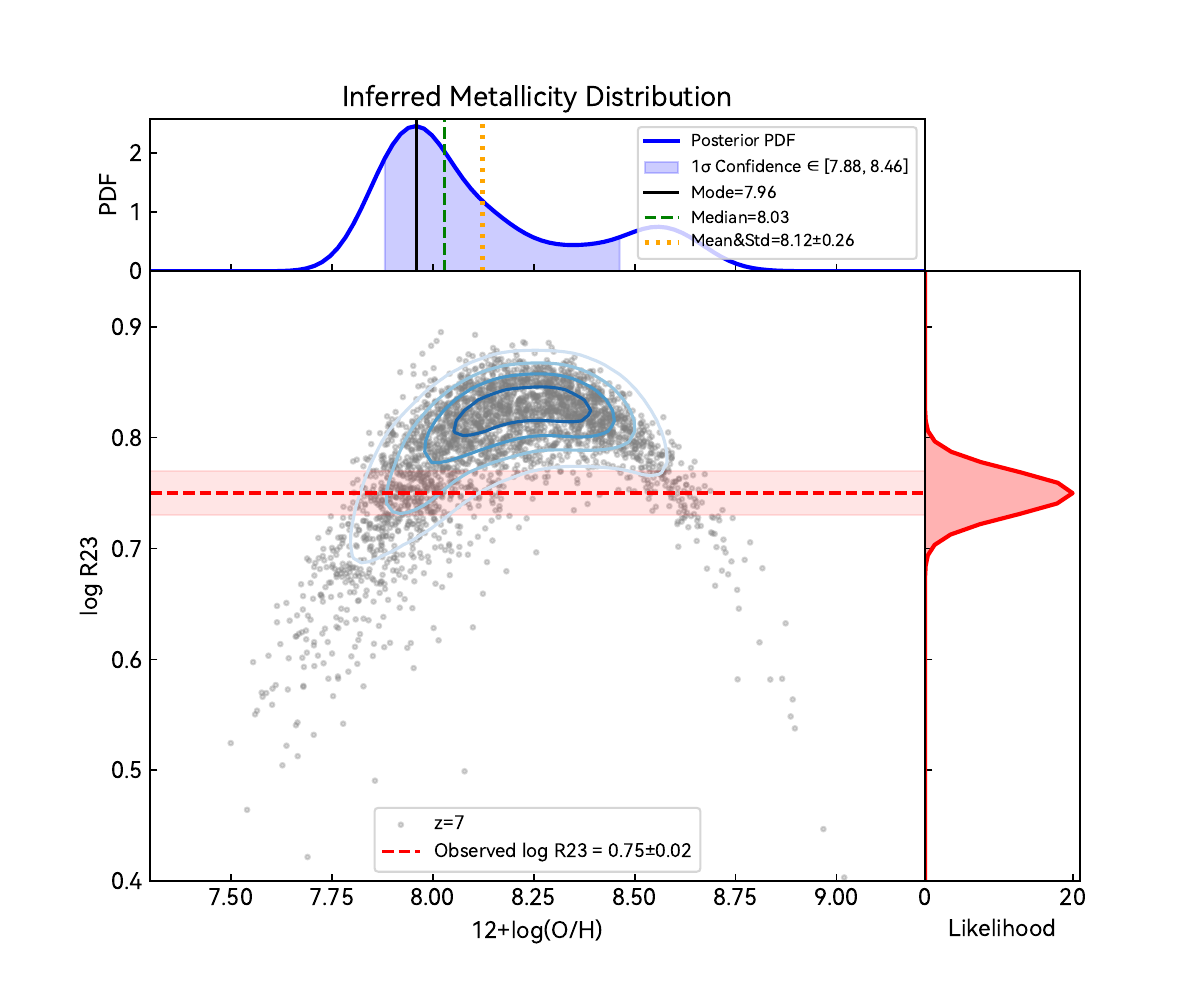}
    \caption{Main panel: The theoretical R23 calibration relation between the line ratio log R23 and $12+\log(\mathrm{O/H})$. Gray points represent the individual photoionization models of our simulated galaxies. Right panel: The assumed observational constraint. We model the measured line ratio (here log R23=0.75) with a Gaussian error distribution ($\sigma = 0.02$ dex, high measurement precision), shown as the red curve. Top panel: The resulting posterior PDF for metallicity, derived via KDE from the intersection of the observed constraint (right panel) with the photoionization model (main panel). Key statistics of the PDF are indicated: the mode (peak), median, mean, standard deviation ($\sigma$), and the 16th/84th percentiles ($\sigma_{16}$ and $\sigma_{84}$) defining the 1$\sigma$ credible interval.}
    \label{fig:kde_example}
\end{figure}

To quantitatively evaluate the performance of different metallicity calibrations under varying observational conditions, we employed a Bayesian framework based on Kernel Density Estimation (KDE). Several Bayesian and model-based approaches have also been developed to infer nebular metallicity and ionization parameter from strong emission lines such as IZI \citep{Blanc2015} and BOND \citep{ValeAsari2016}, allowing full posterior distributions and joint constraints on the derived abundance. Fig. \ref{fig:kde_example} shows an example of a posterior distribution of metallicity from R23 calibration when the observed log R23=0.75$\pm$0.02. The bimodal structure of the PDF directly reflects the dual-branch degeneracy of the R23 diagnostic at this line ratio, where both the metal-rich and metal-poor branches are plausible solutions. This inherent ambiguity, quantified by the broad width and bimodal shape of the PDF, contributes significantly to the total metallicity uncertainty when using this calibration. It is also important to note that our photoionization model sample is uniformly distributed in stellar mass which can not give a strict uniform distribution in log(O/H) due to the dispersion in MZR. Therefore, the KDE effectively uses the metallicity distribution of the selected ASTRID sample as an empirical prior.

We simulate observational noise by assuming a constant absolute flux error ($\sigma_{\mathrm{abs}}$) for all emission lines, a reasonable approximation for spectra dominated by continuum noise. The magnitude of this noise is anchored to the H$\beta$ line, with relative errors ($e_{\mathrm{H}\beta} = \sigma_{\mathrm{abs}} / F_{\mathrm{H}\beta}$) set to 0.01, 0.05, and 0.1, representing high, moderate, and low signal-to-noise ratios (S/N), respectively. Therefore, the relative errors of other emission lines should be:
\begin{equation}
e_{\mathrm{line}}=\frac{\sigma{_\mathrm{abs}}}{F_\mathrm{line}}=e_{\mathrm{H}\beta}\frac{F_{\mathrm{H}\beta}}{F_\mathrm{line}}.
\end{equation}

For single-variable diagnostics (i.e., R23, R3, R2, and N2), the posterior distribution of metallicity depends solely on the observed line ratio and its propagated error. However, for diagnostics involving two independent line ratios (i.e., O3N2, Ne3O2, O32, and N2O2), the analytical error depends on a secondary line ratio (e.g., the error of log O32 depends on itself and log(\oii{}/\hb{}). To address this dimensionality problem and simplify the error analysis, we adopted a binning approach: for a given diagnostic value, we identified the corresponding bin in our models and calculated the median value of the secondary line ratio within that bin. This median value was then used to compute the representative observational error for the primary diagnostic. We performed a test of the uncertainties of metallicity on a grid of two line ratios and find that the impact from the secondary line ratio on the is not significant. Therefore, this dimensionality reduction method will not introduce additional bias.

\subsubsection{Evaluation of calibration performance}

\begin{figure*}
	\includegraphics[width=\textwidth]{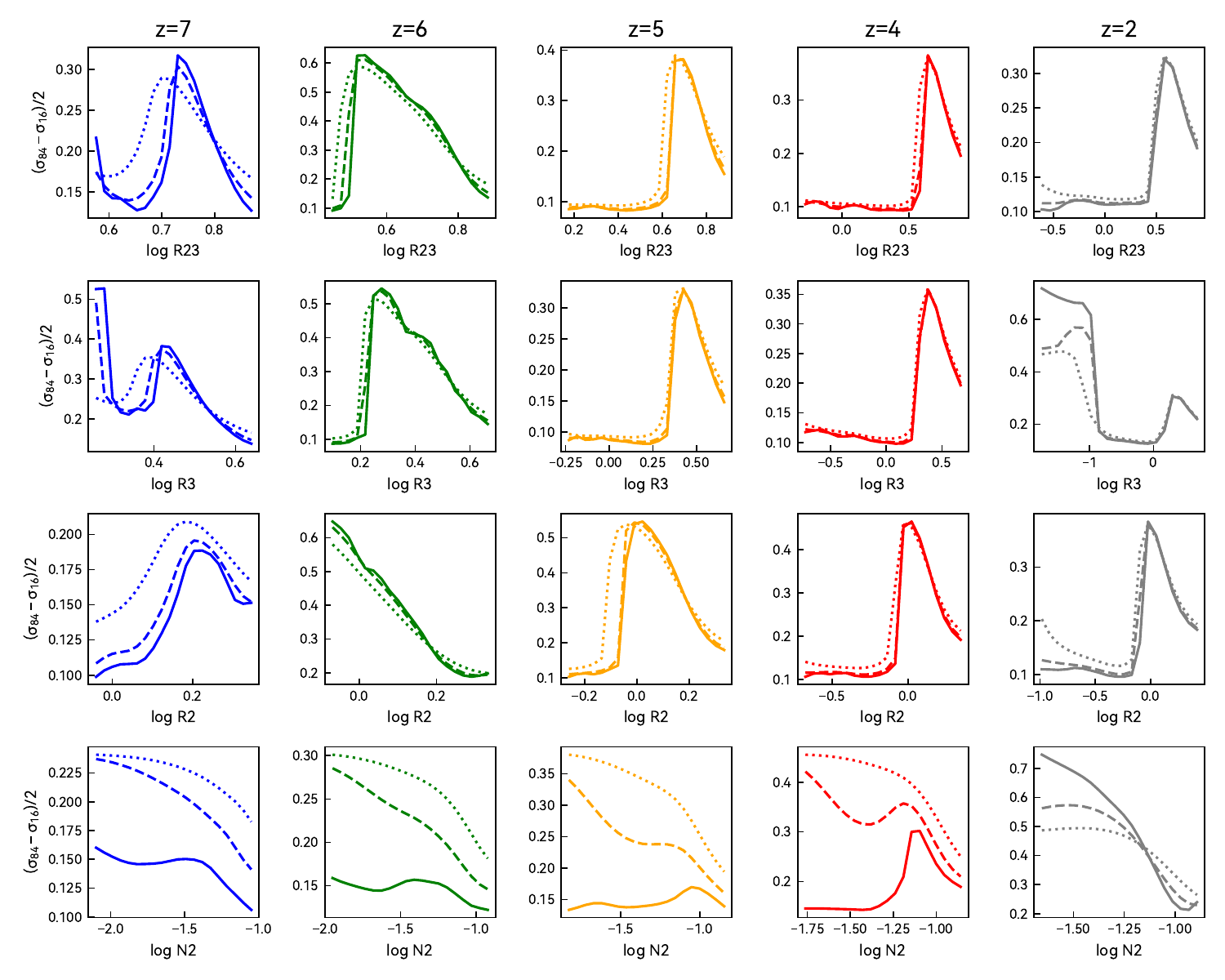}
    \caption{Uncertainty ($\sigma \equiv (\sigma_{84}-\sigma_{16})/2$) in metallicity derived from uni-variate diagnostics (R23, R3, R2, N2) as a function of the observed line ratio. Solid, dashed, and dotted lines correspond to H$\beta$ relative flux errors ($e_{\mathrm{H}\beta}$) of 0.01, 0.05, and 0.1, representing high, moderate, and low S/N regimes, respectively. The prominent peak in uncertainty for log R23 around 0.7-0.8 marks the degenerate turnover region between its upper and lower branches.}
    \label{fig:sigma_1d}
\end{figure*}

\begin{figure*}
	\includegraphics[width=\textwidth]{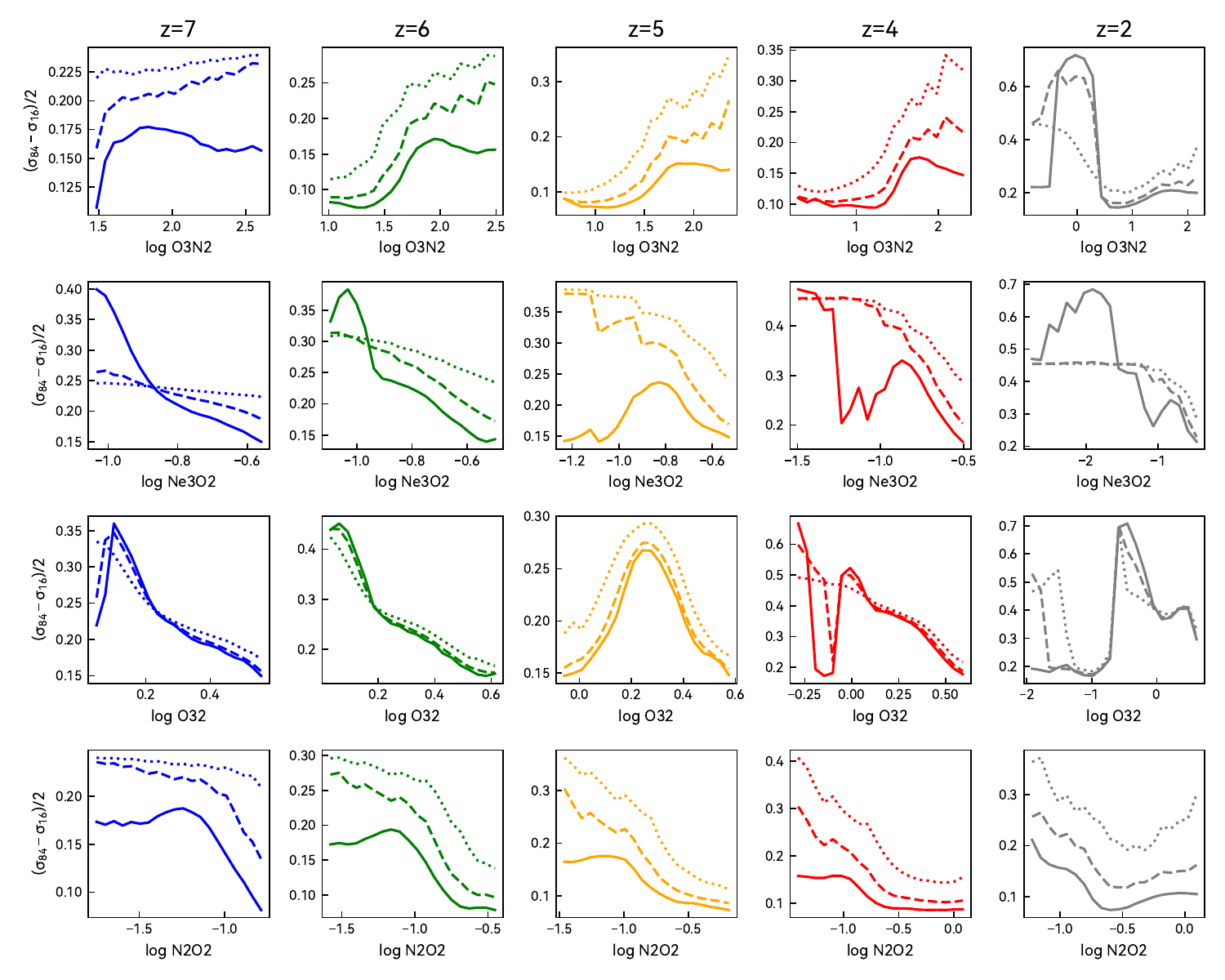}
    \caption{Uncertainty ($\sigma \equiv (\sigma_{84}-\sigma_{16})/2$) in metallicity derived from bi-variate diagnostics (O3N2, Ne3O2, O32, N2O2) as a function of the observed line ratio. The legend is the same as the Fig. \ref{fig:sigma_1d}.}
    \label{fig:sigma_2d}
\end{figure*}

We quantified the uncertainty of the derived metallicities using the 68\% confidence interval of the posterior distribution, defined as $(\sigma_{84} - \sigma_{16}) / 2$. The results for single-variable and two-variable diagnostics as a function of line ratio and S/N are presented in Fig. \ref{fig:sigma_1d} and Fig. \ref{fig:sigma_2d}, respectively. In these figures, solid, dashed, and dotted lines correspond to \hb{} relative errors of 0.01, 0.05, and 0.1. For reference, in the JWST JADES DR4 NIRSpec R1000\_5pix catalog, 0.2\%, 7\%, and 24\% of targets at $z=5-7$ can meet the above high, moderate and low S/N requirements, respectively.

Our analysis yields several key insights regarding the stability and precision of different calibrations:
\begin{enumerate}

\item Common Diagnostics (R23, R3, and R2): The R23 diagnostic emerges as the most robust calibration overall. It is remarkably insensitive to S/N variations, maintaining a dispersion of $(\sigma_{84} - \sigma_{16}) / 2 \simeq 0.1$ dex over most of its dynamic range. However, a significant limitation arises in the ``turnover'' region (log R23 $\simeq$ 0.8), where the upper and lower metallicity branches merge. In this regime, the degeneracy leads to broad PDFs, causing uncertainties to spike to $\sim 0.3$ dex. The R2 and R3 calibrations exhibit similar behavior due to their inherent double-branched nature.

\item Nitrogen-based Diagnostics (N2, O3N2, N2O2): The N2 calibration benefits from a monotonic relation with metallicity, effectively eliminating the branching degeneracy observed in R23. However, it shows a strong sensitivity to S/N. Under high S/N conditions ($e_{\mathrm{H}\beta}=0.01$), the uncertainty is approximately 0.15 dex, but this degrades rapidly to >0.3 dex in low S/N scenarios. This susceptibility is primarily driven by the intrinsic weakness of the \nii{} luminosity. Similar S/N-dependent trends are observed for other nitrogen-bearing diagnostics, such as O3N2 and N2O2. More importantly, nitrogen-based indicators are subject to severe systematic uncertainties arising from the evolution of the N/O ratio. While our Bayesian analysis quantifies the statistical scatter within the ASTRID framework, it cannot account for the $\sim 0.2-0.4$ dex systematic N-enhancement recently observed by JWST in high-redshift star-forming galaxies \citep{Cataldi2026}. If high-redshift galaxies are intrinsically nitrogen-rich compared to the ASTRID model, applying our N-based calibrations would lead to a significant overestimation of the gas-phase oxygen abundance. Furthermore, given the stochastic nature of nitrogen enrichment driven by bursty star formation \citep{McClymont2026}, a universal $N$-based calibration may be inherently unreliable at $z > 2$. Therefore, we emphasize that while these diagnostics are useful for breaking the R23 degeneracy in high-quality spectra, their results should be cross-checked with $\alpha$-element tracers (e.g., Ne3O2 or R23) to mitigate chemical evolution biases.

\item Ionization-sensitive Diagnostics (O32, Ne3O2): Calibrations based on ionization parameter proxies, specifically O32 and Ne3O2, generally exhibit larger uncertainties compared to R23 or N2. We observe a redshift dependence, where errors are more pronounced at lower redshifts ($z \lesssim 4$). Their performance improves marginally at high redshifts, particularly for high values of O32 and Ne3O2, where the uncertainty can be constrained to $\lesssim 0.2$ dex.

\end{enumerate}

Based on these findings, we propose a hierarchy of calibration choices depending on spectral quality and the physical regime of the target galaxy. For spectra with low S/N, R23 is the preferred diagnostic due to its high precision and relative insensitivity to noise, provided the galaxy does not lie within the turnover region. In cases where spectral coverage is limited, R3 or R2 serve as viable single-line alternatives. Conversely, for high-quality spectra (high S/N and resolution), if the log R23 value falls within the ambiguous turnover region ($0.7-0.8$) or if the branch cannot be determined via other means, N2O2, O3N2 or N2 are excellent alternatives to break the degeneracy, despite their intrinsic scatter. A modified R23 calibration that involve N2O2 and O32 as parameters can be an alternative approach \citep[e.g., $\rm R_u$ in ][]{Liu2026}. It should be noted that nitrogen-based calibrations may introduce additional systematic uncertainties due to potential variations in the N/O ratio, and N2O2 is significantly affected by dust extinction. Finally, while O32 and Ne3O2 generally yield larger uncertainties, they remain useful in specific niches and will not be affected by the non-$\alpha$ enrichment. We recommend their use primarily for high-redshift, high-ionization sources. Additionally, Ne3O2 is particularly valuable in regions with uncertain dust attenuation corrections, owing to the close wavelength proximity of the \neiii{}$\lambda$3869 and \oii{}$\lambda$3727 lines. For calibrations at $z\leq2$ , we still recommend using the calibrations fully based on $T_e$ \citep[e.g.,][]{Andrews2013,Curti2020} as the direct-$T_e$ method is well-established and almost independent of the model.

\section{Summary} \label{sec:summary}

In this work, we present a theoretical framework for deriving redshift-dependent metallicity calibrations at $z=2-7$ by combining the ASTRID cosmological simulation with \texttt{MAPPINGS V} photoionization modeling. Our main conclusions are summarized as follows:

(i) \textbf{Model Methodology and Validation:} We have developed a theoretical framework that self-consistently computes \hii{} region properties—including ionization parameter, gas pressure, and density-by coupling stellar feedback with the local ISM environment. Our model successfully reproduces the SFR indicators of \cite{Kennicutt1998}, yielding conversion factors ($C_{\rm H\alpha}\simeq41.357$, $C_{\rm [O~II]}\simeq41.015$) in excellent agreement with established literature values. In addition, the predicted \oiii{}$\lambda$5007 luminosity functions across $z=2-7$ are broadly consistent with recent JWST observations, supporting the physical realism of our nebular emission modeling.

(ii) \textbf{Performance of Metallicity Diagnostics:} Through a Bayesian analysis of calibration uncertainties, we find that the R23 diagnostic remains the most robust optical indicator overall. It exhibits high precision and minimal sensitivity to S/N variations across most of its dynamic range. In contrast, other traditional diagnostics exhibit various limitations: ionization-sensitive ratios (e.g., O32, Ne3O2) show large scatter, while UV indicators (e.g., C3O3, N3O3) currently suffer from substantial intrinsic dispersion and complex high-ionization physics. Nitrogen-based diagnostics (e.g., N2, O3N2), although monotonic with metallicity, show systematic offsets at high metallicities and are highly sensitive to observational noise. Dust extinction correction is still necessary when using calibrations with large wavelength differences, such as N2O2.

(iii) \textbf{The Nitrogen Challenge and Future Outlook:} The ASTRID simulation faces challenges in robustly tracking nitrogen and carbon abundances at high redshift. The simulated N/O-O/H relation does not reproduce either the local scaling relations and the systematic $0.18-0.4$ dex nitrogen enhancement recently observed by JWST in star-forming galaxies. These discrepancies likely arise from uncertainties in AGB stellar yield tables and the limited ability of the simulation to capture bursty star-formation and short timescale fluctuations in chemical enrichment.

In conclusion, R23 remains a reliable tool for interpreting current JWST spectroscopic data. However, the development of more accurate high-redshift metallicity diagnostics will require improved modeling of nitrogen and carbon enrichment in cosmological simulations. Future cosmological simulations must incorporate more sophisticated yield prescriptions and better resolve the stochastic, feedback-driven nature of chemical enrichment in the early Universe. Until such improvements are achieved, metallicity estimates derived solely from nitrogen- or carbon-based diagnostics at high redshift should be cross-validated with $\alpha$-element tracers to ensure robust results.

\section*{Acknowledgments}

This project is supported by the Australian Research Council through the Discovery Program (project number DP240101472) funded by the Australian Government. 

\section*{Data Availability}
 
The original simulation data of this article can be accessed at \url{https://astrid-portal.psc.edu/simulation/1/}. The code to reproduce the simulation is available at \url{https://github.com/MP-Gadget/MP-Gadget}, and continues to be developed. FITS forms of the data presented here and the scripts to generate our tables and figures are also available upon request.



\bibliographystyle{mnras}
\bibliography{example} 




\appendix

\section{The variation of parameters in particle splitting} \label{apdx:split}

We select two test galaxies (31.1 and 47.268) to test the stability of line ratios under different parameters ($M_{\rm min}$, $M_{\rm max}$, and $\beta$) of the cluster function. The results are listed in Table \ref{tab:vary_m_beta} and \ref{tab:vary_m_beta_47}. The bolded rows represent the parameters adopted in our research. We can see that the variations of the line ratios are < 0.1 dex.

\begin{table}
    \centering
    \caption{The line ratios of 31.1 under different $M_{\rm min}$, $M_{\rm max}$, and $\beta$}
    \label{tab:vary_m_beta}
    \begin{tabular}{cccccc}
        \hline
        Mass range ($M_\odot$) & $\beta$ & log R3 & log R2 & log N2 \\
        \hline
        \multirow{3}{*}{$1\times10^3-2\times10^5$} &1.8 & 0.419 & 0.234 & -1.031 \\
        & 2.0 & 0.397 & 0.237 & -1.027 \\
        & 2.2 & 0.369 & 0.251 & -1.023 \\
        \hline
        \multirow{3}{*}{$2\times10^3-4\times10^5$} & 1.8 & 0.444 & 0.224 & -1.036 \\
        & \bf{2.0} & \bf{0.437} & \bf{0.221} & \bf{-1.033} \\
        & 2.2 & 0.397 & 0.240 & -1.028 \\
        \hline
        \multirow{3}{*}{$4\times10^3-8\times10^5$} & 1.8 & 0.461 & 0.215 & -1.039 \\
        & 2.0 & 0.448 & 0.222 & -1.036 \\
        & 2.2 & 0.424 & 0.229 & -1.033 \\
        \hline
    \end{tabular}
\end{table}

\begin{table}
    \centering
    \caption{The line ratios of 47.268 under different $M_{\rm min}$, $M_{\rm max}$, and $\beta$}
    \label{tab:vary_m_beta_47}
    \begin{tabular}{cccccc}
        \hline
        Mass range ($M_\odot$) & $\beta$ & log R3 & log R2 & log N2 \\
        \hline
        \multirow{3}{*}{$1\times10^3-2\times10^5$} & 1.8 & 0.518 & 0.226 & -1.248 \\
        & 2.0 & 0.491 & 0.252 & -1.220 \\
        & 2.2 & 0.484 & 0.263 & -1.205 \\
        \hline
        \multirow{3}{*}{$2\times10^3-4\times10^5$} & 1.8 & 0.540 & 0.198 & -1.277 \\
        & \bf{2.0} & \bf{0.522} & \bf{0.217} & \bf{-1.257} \\
        & 2.2 & 0.503 & 0.240 & -1.234 \\
        \hline
        \multirow{3}{*}{$4\times10^3-8\times10^5$} & 1.8 & 0.555 & 0.176 & -1.300 \\
        & 2.0 & 0.545 & 0.191 & -1.284 \\
        & 2.2 & 0.527 & 0.211 & -1.264 \\
        \hline
    \end{tabular}
\end{table}

\section{The coefficient of calibrations} \label{apdx:coefficient}

Here we list the coefficients of optical calibrations in Table \ref{tab:optical_table} and the coefficients of UV calibrations in Table \ref{tab:uv_table}. The digital form of this table can be accessed at \url{https://github.com/pentyum/astrid_metallicity/}. The metallicity ranges of our fitting at $z$=7,6,5,4,2 are $7.75-8.60$, $7.80-8.81$, $7.81-8.99$, $7.81-9.14$, $7.77-9.16$, respectively. It is important to notice that due to the large coefficients in some calibrations, extrapolated values outside the fitted range can be invalid and should not be used. The fits are dust-free. Observed fluxes should be corrected for attenuation before using ratios with widely separated wavelengths, especially for N2O2.

\begin{table*}
	\centering
	\caption{The coefficient of optical calibrations ($y=a_0+a_1x+a_2x^2+a_3x^3$, $x$=12+log(O/H), $y$=logarithm of the line ratio)}
	\label{tab:optical_table}
	\begin{tabular}{cccccc} 
		\hline
		  Calibrator & Redshift & $a_0$ & $a_1$ & $a_2$ & $a_3$ \\
		\hline
\multirow{5}{*}{R23} & 7 & 17.102$\pm$12.056 & -10.964$\pm$4.456 & 1.938$\pm$0.549 & -0.103$\pm$0.023 \\
 & 6 & 27.859$\pm$7.066 & -15.453$\pm$2.587 & 2.556$\pm$0.315 & -0.131$\pm$0.013 \\
 & 5 & 72.909$\pm$6.526 & -32.183$\pm$2.361 & 4.628$\pm$0.284 & -0.217$\pm$0.011 \\
 & 4 & 77.314$\pm$6.810 & -34.316$\pm$2.450 & 4.953$\pm$0.293 & -0.232$\pm$0.012 \\
 & 2 & 91.965$\pm$5.581 & -40.878$\pm$2.023 & 5.912$\pm$0.244 & -0.278$\pm$0.010 \\
\hline
\multirow{5}{*}{R3} & 7 & 22.941$\pm$17.546 & -13.383$\pm$6.486 & 2.262$\pm$0.799 & -0.117$\pm$0.033 \\
 & 6 & 28.697$\pm$10.131 & -16.254$\pm$3.709 & 2.711$\pm$0.452 & -0.140$\pm$0.018 \\
 & 5 & 87.072$\pm$9.290 & -37.807$\pm$3.360 & 5.364$\pm$0.405 & -0.249$\pm$0.016 \\
 & 4 & 109.913$\pm$10.186 & -46.790$\pm$3.664 & 6.533$\pm$0.439 & -0.299$\pm$0.017 \\
 & 2 & 118.301$\pm$9.968 & -50.951$\pm$3.613 & 7.187$\pm$0.436 & -0.332$\pm$0.018 \\
\hline
\multirow{5}{*}{R2} & 7 & 28.785$\pm$17.912 & -15.250$\pm$6.621 & 2.412$\pm$0.815 & -0.119$\pm$0.033 \\
 & 6 & 95.163$\pm$12.576 & -39.515$\pm$4.604 & 5.365$\pm$0.561 & -0.239$\pm$0.023 \\
 & 5 & 115.293$\pm$11.838 & -47.426$\pm$4.282 & 6.399$\pm$0.516 & -0.284$\pm$0.021 \\
 & 4 & 89.011$\pm$10.649 & -37.666$\pm$3.830 & 5.201$\pm$0.459 & -0.235$\pm$0.018 \\
 & 2 & 91.540$\pm$6.897 & -40.534$\pm$2.500 & 5.797$\pm$0.302 & -0.270$\pm$0.012 \\
\hline
\multirow{5}{*}{O3N2} & 7 & 46.950$\pm$61.528 & -19.206$\pm$22.743 & 2.787$\pm$2.801 & -0.136$\pm$0.115 \\
 & 6 & -5.358$\pm$34.942 & -1.242$\pm$12.791 & 0.736$\pm$1.560 & -0.058$\pm$0.063 \\
 & 5 & 52.209$\pm$29.371 & -20.734$\pm$10.625 & 2.919$\pm$1.280 & -0.139$\pm$0.051 \\
 & 4 & 51.441$\pm$24.539 & -20.976$\pm$8.827 & 3.000$\pm$1.057 & -0.144$\pm$0.042 \\
 & 2 & -86.395$\pm$16.188 & 28.890$\pm$5.868 & -3.018$\pm$0.708 & 0.098$\pm$0.028 \\
\hline
\multirow{5}{*}{Ne3O2} & 7 & -12.311$\pm$28.134 & 3.593$\pm$10.399 & -0.335$\pm$1.281 & 0.008$\pm$0.053 \\
 & 6 & -57.424$\pm$18.458 & 19.344$\pm$6.757 & -2.153$\pm$0.824 & 0.078$\pm$0.033 \\
 & 5 & -15.302$\pm$16.712 & 4.182$\pm$6.045 & -0.335$\pm$0.728 & 0.005$\pm$0.029 \\
 & 4 & 37.339$\pm$16.483 & -15.954$\pm$5.929 & 2.215$\pm$0.710 & -0.102$\pm$0.028 \\
 & 2 & 54.099$\pm$13.209 & -21.471$\pm$4.788 & 2.824$\pm$0.578 & -0.124$\pm$0.023 \\
\hline
\multirow{5}{*}{O32} & 7 & -5.855$\pm$30.508 & 1.871$\pm$11.277 & -0.151$\pm$1.389 & 0.002$\pm$0.057 \\
 & 6 & -66.468$\pm$19.544 & 23.262$\pm$7.154 & -2.654$\pm$0.872 & 0.099$\pm$0.035 \\
 & 5 & -28.217$\pm$17.742 & 9.617$\pm$6.418 & -1.035$\pm$0.773 & 0.035$\pm$0.031 \\
 & 4 & 20.908$\pm$17.230 & -9.126$\pm$6.198 & 1.331$\pm$0.742 & -0.064$\pm$0.030 \\
 & 2 & 26.759$\pm$13.734 & -10.416$\pm$4.978 & 1.390$\pm$0.601 & -0.062$\pm$0.024 \\
\hline
\multirow{5}{*}{N2} & 7 & -23.998$\pm$53.074 & 5.819$\pm$19.618 & -0.525$\pm$2.416 & 0.018$\pm$0.099 \\
 & 6 & 34.057$\pm$30.539 & -15.013$\pm$11.180 & 1.975$\pm$1.363 & -0.082$\pm$0.055 \\
 & 5 & 34.866$\pm$24.834 & -17.074$\pm$8.983 & 2.445$\pm$1.082 & -0.110$\pm$0.043 \\
 & 4 & 58.468$\pm$18.563 & -25.812$\pm$6.677 & 3.532$\pm$0.800 & -0.155$\pm$0.032 \\
 & 2 & 204.700$\pm$9.188 & -79.842$\pm$3.330 & 10.205$\pm$0.402 & -0.430$\pm$0.016 \\
\hline
\multirow{5}{*}{N2O2} & 7 & -53.121$\pm$54.116 & 21.503$\pm$20.003 & -3.006$\pm$2.463 & 0.141$\pm$0.101 \\
 & 6 & -60.818$\pm$31.809 & 24.714$\pm$11.644 & -3.432$\pm$1.420 & 0.159$\pm$0.058 \\
 & 5 & -79.747$\pm$26.282 & 30.428$\pm$9.507 & -3.981$\pm$1.145 & 0.176$\pm$0.046 \\
 & 4 & -28.081$\pm$18.487 & 11.289$\pm$6.650 & -1.619$\pm$0.796 & 0.079$\pm$0.032 \\
 & 2 & 113.155$\pm$7.732 & -38.946$\pm$2.803 & 4.341$\pm$0.338 & -0.157$\pm$0.014 \\
		\hline
	\end{tabular}
\end{table*}

\begin{table*}
	\centering
	\caption{The coefficient of UV calibrations ($y=a_0+a_1x+a_2x^2+a_3x^3$, $x$=12+log(O/H), $y$=logarithm of the line ratio)}
	\label{tab:uv_table}
	\begin{tabular}{cccccc} 
		\hline
		  Calibrator & Redshift & $a_0$ & $a_1$ & $a_2$ & $a_3$ \\
		\hline
\multirow{5}{*}{C3O3} & 7 & -85.785$\pm$12.617 & 32.070$\pm$4.664 & -4.009$\pm$0.574 & 0.168$\pm$0.024 \\
 & 6 & -110.199$\pm$9.685 & 41.640$\pm$3.545 & -5.256$\pm$0.432 & 0.222$\pm$0.018 \\
 & 5 & -148.142$\pm$12.803 & 55.944$\pm$4.633 & -7.054$\pm$0.558 & 0.298$\pm$0.022 \\
 & 4 & -146.456$\pm$15.529 & 55.156$\pm$5.595 & -6.943$\pm$0.671 & 0.293$\pm$0.027 \\
 & 2 & 197.664$\pm$17.131 & -70.044$\pm$6.234 & 8.221$\pm$0.755 & -0.318$\pm$0.030 \\
\hline
\multirow{5}{*}{N3O3} & 7 & -62.627$\pm$49.525 & 23.124$\pm$18.306 & -2.930$\pm$2.254 & 0.125$\pm$0.092 \\
 & 6 & -88.742$\pm$30.259 & 33.344$\pm$11.077 & -4.245$\pm$1.351 & 0.181$\pm$0.055 \\
 & 5 & -180.825$\pm$25.466 & 66.261$\pm$9.216 & -8.165$\pm$1.111 & 0.337$\pm$0.045 \\
 & 4 & -108.867$\pm$21.453 & 39.804$\pm$7.729 & -4.926$\pm$0.927 & 0.205$\pm$0.037 \\
 & 2 & 171.999$\pm$15.760 & -61.835$\pm$5.735 & 7.331$\pm$0.695 & -0.287$\pm$0.028 \\
\hline
\multirow{5}{*}{N2O2 UV} & 7 & 25.776$\pm$78.053 & -12.977$\pm$28.851 & 1.933$\pm$3.553 & -0.092$\pm$0.146 \\
 & 6 & 21.829$\pm$42.509 & -10.694$\pm$15.561 & 1.573$\pm$1.897 & -0.074$\pm$0.077 \\
 & 5 & 32.961$\pm$32.036 & -15.956$\pm$11.589 & 2.361$\pm$1.396 & -0.112$\pm$0.056 \\
 & 4 & 222.051$\pm$23.434 & -85.025$\pm$8.429 & 10.764$\pm$1.010 & -0.453$\pm$0.040 \\
 & 2 & 300.183$\pm$10.606 & -113.200$\pm$3.845 & 14.180$\pm$0.464 & -0.592$\pm$0.019 \\
\hline
\multirow{5}{*}{O32 UV} & 7 & 165.039$\pm$35.545 & -61.435$\pm$13.139 & 7.715$\pm$1.618 & -0.326$\pm$0.066 \\
 & 6 & 137.620$\pm$27.196 & -52.790$\pm$9.956 & 6.848$\pm$1.214 & -0.299$\pm$0.049 \\
 & 5 & 171.339$\pm$30.332 & -65.488$\pm$10.977 & 8.448$\pm$1.323 & -0.367$\pm$0.053 \\
 & 4 & 136.699$\pm$31.372 & -52.872$\pm$11.303 & 6.916$\pm$1.356 & -0.305$\pm$0.054 \\
 & 2 & -284.816$\pm$26.274 & 102.627$\pm$9.561 & -12.153$\pm$1.158 & 0.473$\pm$0.047 \\
\hline
\multirow{5}{*}{C32} & 7 & 31.688$\pm$27.065 & -11.589$\pm$10.004 & 1.497$\pm$1.232 & -0.066$\pm$0.051 \\
 & 6 & -10.861$\pm$18.806 & 3.230$\pm$6.884 & -0.205$\pm$0.839 & -0.002$\pm$0.034 \\
 & 5 & -13.911$\pm$18.489 & 4.481$\pm$6.688 & -0.369$\pm$0.806 & 0.005$\pm$0.032 \\
 & 4 & 1.646$\pm$17.914 & -1.490$\pm$6.444 & 0.389$\pm$0.772 & -0.027$\pm$0.031 \\
 & 2 & -142.262$\pm$13.403 & 51.894$\pm$4.858 & -6.188$\pm$0.586 & 0.242$\pm$0.024 \\
\hline
\multirow{5}{*}{N32} & 7 & 78.531$\pm$48.815 & -26.006$\pm$18.044 & 2.938$\pm$2.222 & -0.113$\pm$0.091 \\
 & 6 & 32.075$\pm$27.085 & -10.595$\pm$9.915 & 1.261$\pm$1.209 & -0.053$\pm$0.049 \\
 & 5 & 16.121$\pm$23.995 & -4.738$\pm$8.680 & 0.548$\pm$1.046 & -0.025$\pm$0.042 \\
 & 4 & -29.197$\pm$20.179 & 11.666$\pm$7.258 & -1.434$\pm$0.869 & 0.055$\pm$0.035 \\
 & 2 & -249.705$\pm$14.391 & 92.845$\pm$5.216 & -11.374$\pm$0.630 & 0.460$\pm$0.025 \\
\hline
\multirow{5}{*}{C23He2} & 7 & -108.231$\pm$126.806 & 36.661$\pm$46.892 & -4.007$\pm$5.777 & 0.144$\pm$0.237 \\
 & 6 & -500.860$\pm$65.234 & 181.467$\pm$23.887 & -21.760$\pm$2.914 & 0.867$\pm$0.118 \\
 & 5 & -324.594$\pm$57.957 & 119.046$\pm$20.999 & -14.395$\pm$2.534 & 0.578$\pm$0.102 \\
 & 4 & -228.576$\pm$58.095 & 83.986$\pm$20.956 & -10.132$\pm$2.517 & 0.405$\pm$0.101 \\
 & 2 & -433.828$\pm$32.171 & 159.859$\pm$11.756 & -19.414$\pm$1.431 & 0.781$\pm$0.058 \\
\hline
\multirow{5}{*}{O3He2} & 7 & 0.768$\pm$130.918 & -4.343$\pm$48.413 & 1.151$\pm$5.964 & -0.074$\pm$0.245 \\
 & 6 & -350.551$\pm$68.545 & 124.416$\pm$25.099 & -14.527$\pm$3.061 & 0.560$\pm$0.124 \\
 & 5 & -199.904$\pm$61.673 & 70.960$\pm$22.349 & -8.200$\pm$2.697 & 0.310$\pm$0.108 \\
 & 4 & -139.729$\pm$64.554 & 49.200$\pm$23.295 & -5.574$\pm$2.799 & 0.205$\pm$0.112 \\
 & 2 & -553.346$\pm$40.034 & 199.984$\pm$14.644 & -23.812$\pm$1.784 & 0.936$\pm$0.072 \\
		\hline
	\end{tabular}
\end{table*}


\bsp	
\label{lastpage}
\end{document}